# "Forbidden" polarisation and extraordinary piezoelectric effect in organometallic lead halide perovskites

Milica Vasiljevic[1], Márton Kollár[4], David Spirito[3], Lukas Riemer[1], László Forró[2,5], Endre Horváth[2,4,6], Semën Gorfman[3], Dragan Damjanovic[1,*]

1. Group for Ferroelectrics and Functional Oxides, Ecole Polytechnique Fédérale de Lausanne, 1015 Lausanne, Switzerland
2. Laboratory of Physics of Complex Matter, Ecole Polytechnique Fédérale de Lausanne, 1015 Lausanne, Switzerland
3. The Department of Materials Science and Engineering, Tel Aviv University, Ramat Aviv, Tel Aviv 6997801, Israel
4. KEP Technologies, 1228 Plan-les-Ouates, Switzerland
5. Stavropoulos Center for Complex Quantum Matter, University of Notre Dame, IN 46556, USA
6. Laboratory for Quantum Magnetism, École Polytechnique Fédérale de Lausanne, 1015 Lausanne, Switzerland

*email: dragan.damjanovic@epfl.ch

Abstract

Organometallic lead halide perovskites are highly efficient materials for solar cells and other optoelectronic applications due to their high quantum efficiency and exceptional semiconducting properties. A peculiarity of these perovskites is the substantial ionic motion under external forces. Here, we reveal that electric field-and light-induced ionic motion in $MAPbX_3$ crystals (X=Cl, Br, I and MA=$CH_3NH_3$) leads to unexpected piezoelectric-like response, an order of magnitude larger than in ferroelectric perovskite oxides. The nominal macroscopic symmetry of the crystals is broken by redistribution of ionic species, which can be controlled deterministically by light and electric field. The revealed piezoelectric response is possibly present in other materials with significant ionic activity but the unique feature of organometallic perovskites is the strong effect on the piezoelectric response of interplay of ionic motion ($MA^+$ and $X^{-1}$) and photoelectrons generated with illumination.

Organometallic lead halide perovskites, represented by methylammonium (MA) lead halides MAPbX$_3$ (X=Cl, Br, I), exhibit unusually broad spectrum of properties attractive for optoelectronic applications such as solar cells, radiation detectors, light-emitting diodes, and photodetectors[1]. Despite extensive studies, some key aspects of MAPbX$_3$ behaviour are still poorly understood. Among the most intriguing are the nature of ionic defects, their interaction with external fields and the role of their transport on properties.[2] MA$^{+1}$, Pb$^{+2}$, X$^{-1}$ and H$^{+1}$ are directly or indirectly known to form vacancies and possibly interstitials.[3,4,5] The charge-compensating species (electrons, holes or other ionic defects) for each defect are most often not identified. There is substantial evidence that ionic defects in MAPbX$_3$ are much more mobile at ambient temperature[6,7,8] than their counterparts in oxide perovskites.[9] Additionally, the MA$^{+1}$ molecule (with an equivalent radius of ≈2.17 Å[10] it is comparable in size with X$^{-1}$ and Pb$^{+2}$ ions[11]) and defect associates (clusters that ionic defects form with their charge-compensating species) may reorient in external electric field. Since charged point defects have different size than the host site, they exert strain on their lattice surrounding deforming the unit cell and distorting halide octahedra, X$_6$. Various orientations of MA$^{+1}$ and defect clusters also deform the lattice. The charged ionic defects, orientable molecules and defect clusters therefore interact with both electric field and stress, i.e. they are electro-mechanically active.[12] Potential for a strong electro-mechanical effect in these materials is significant: for example, a high concentration of charged iodine vacancies in apical position of I$_6$ octahedra can contract the unit cell by 11%.[8] Three electro-mechanical effects are usually considered in deformable, polarizable materials. Electrostriction, which is present in all materials regardless of their symmetry, has been reported to be large in MAPbI$_3$ when compared to oxide perovskites.[13] Flexoelectric effect may also appear in all materials, but requires stress (strain) or polarization (electric field) gradients and is thus usually negligible in bulk samples. Strong light-assisted flexoelectric effect has been measured in thin sheets of MAPbX$_3$ for X=Cl and Br.[14] Piezoelectricity is the most interesting electro-mechanical effect for applications and is the largest of the three in a given bulk material. Importantly, piezoelectricity is forbidden in centrosymmetric crystals.[15] The nominal average crystal structure of the bromide and chloride is considered cubic centrosymmetric, space group type *Pm-3m,* at room temperature[16] and thus does not support piezoelectricity. The symmetry of the iodide at room temperature is usually given as centrosymmetric tetragonal *I*4/*mcm.* Therefore, even if individual defects and charged molecules in an MAPbX$_3$ crystal are electro-mechanically active, the sample will not be piezoelectric if the arrangement of defects and orientation of molecules is random. Ferroelectric-like polar behaviour has been reported for MAPbI$_3$,[17,18] in which case the proposed symmetry is *I*4*cm* and the material is potentially piezoelectric.[17,19,20] Macroscopic polarity in MAPbI$_3$ is likely related to ordering of MA$^{+1}$ molecules.[21] It has been inconclusively speculated that polarization may enhance optoelectronic properties.[22]

We show here that bulk samples of MAPbX$_3$ single crystals exhibit macroscopic polarity and an enormous apparent converse piezoelectric-like response at low frequencies. The presented evidence points toward ionic rearrangement as the main cause of the symmetry breaking and ensuing polarity and piezoelectricity. The "forbidden" polarization and piezoelectricity can arise spontaneously, during the crystal growth, and may be triggered and controlled by electric field and UV light. The results imply that the electrically and optically generated piezoelectric strain may affect operation of optoelectronic devices.[23,24,25]

**Results**

Crystals were prepared by the inverse crystallization method (see Methods).[26] Electrodes were deposited on two opposite faces of as-received crystals shaped as cuboids (X=Cl, Br) or irregular polyhedra (X=I), Figure 1a. Electric field of variable frequency and strength was applied on the electrodes and mechanical displacement was measured in longitudinal mode with a linear variable differential transformer (LVDT). The current through the samples is measured by connecting a resistor in series with the sample and measuring voltage drop across the resistor (see Methods, and Supplementary Note 1 and 2).

**Induced piezoelectricity**

Figure 1b-d illustrates mechanical displacement generated in crystals upon application of electric field. In all sufficiently insulating materials an alternating electric field $E_{ac} = E_0 \sin(\omega t)$ induces electrostrictive strain $x = ME_{ac}^2$ where M is the electrostrictive coefficient.[27] The piezoelectric effect may be induced in centrosymmetric materials if a direct electric-field bias $E_{dc}$ is applied together with the $E_{ac}$.[28] The role of the $E_{dc}$ is to break the crystal's centric symmetry and bias electrostriction, similar to what polarization does in ferroelectrics. The total strain $x$ is then given by $x = M(E_{ac} + E_{dc})^2$. The piezoelectric (linear) component of the strain is then $x = (2ME_{dc})E_{ac} = dE_{ac}$, where $d = 2ME_{dc}$ is the apparent piezoelectric coefficient (see Supplementary Note 3). This simultaneous appearance of the electrostrictive and piezoelectric strains is illustrated in Figure 1b, which shows displacement of an MAPbCl$_3$ crystal on which an alternating field with frequency of 10 mHz, amplitude ≈230 V/cm and an offset of ≈ -34 V/cm was applied. The electrostrictive strain is identified in the graph as the second harmonic response (twice the frequency of the applied field) whereas the piezoelectric strain appears as the first harmonic response. In a general case, electric-field induced alternating strain in a material that exhibits both piezoelectric and electrostrictive effects is given by equation $x = ME_{ac}^2 + dE_{ac}$ (see Supplementary Note 3). The fit (purple dots) of the data (blue line) in Figure 1b with this equation leads to M≈1.45×10$^{-12}$ m$^2$/V$^2$ and d≈16,900 pm/V. For comparison, the latter value is 30-80 times larger than in Pb(Zr,Ti)O$_3$ ceramics (d≈200-

500 pm/V) and up to ten times larger than in relaxor ferroelectric Pb(Mg$_{1/3}$Nb$_{2/3}$)O$_3$-PbTiO$_3$ single crystals (d≈1500-2000 pm/V)[29]. Those oxide perovskites are presently the most widely used piezoelectric materials. The extracted electrostrictive coefficient M for the examined MAPbCl$_3$ and MAPbBr$_3$ samples is several orders of magnitude larger than in a broad class of materials, including MAPbI$_3$[13], as shown in Supplementary Note 4.

To further confirm the piezoelectric nature of the strain signal, Figure 1c illustrates inversion of the strain phase angle in MAPbBr$_3$ when the sign of E$_{dc}$ is switched, as expected from relation $x = (2ME_{dc})E_{ac}$. In contrast to the data for the chloride in Figure 1b, only the first harmonic is observed in this bromide sample. Because of a large E$_{dc}$ (in this experiment, V$_{dc}$=±50 V, V$_{ac}$=50 V) the piezoelectric term dominates the total strain and the electrostrictive component of the strain is not readily apparent. The value of d in bromide crystals is also exceptionally large, around 3000 pm/V, and values over 9000 pm/V have been observed in some samples (Supplementary Note 5). Figure 1d shows displacement-voltage relationship for an MAPbI$_3$ sample. The equivalent piezoelectric coefficient in this sample is about 350 pm/V, much larger than previously reported using local techniques (~5 to 25 pm/V)[17] and direct piezoelectric measurements (~50 pm/V)[20]. The asymmetry of the displacement-voltage graph for the positive and the negative V$_{dc}$, indicates asymmetry of the sample.

**Evidence of internal asymmetry**

The apparent piezoelectric coefficient obtained from $x = ME_{ac}^2 + dE_{ac}$ was compared with the piezoelectric coefficient expected from d= $2ME_{dc}$, using the value of M extracted from the former equation. We get discrepancy that, generally, depends irregularly on E$_{dc}$ (Supplementary Note 6). In the case of MAPbCl$_3$ in Figure 1b, the estimated d is ~10,000 pm/V, which is just over half of the value calculated from the experimental data. This discrepancy suggests that the sample has an internal bias field in addition to experiencing the external E$_{dc}$. Another evidence of built-in asymmetry is the piezoelectric displacement generated by an E$_{ac}$ signal (no E$_{dc}$ offset) applied on an as-grown MAPbBr$_3$ crystal, Figure 2a. Combination of the electrostrictive and piezoelectric response is clearly seen, which is inconsistent with the nominal *Pm*-3*m* structure, where only electrostrictive response is expected. The displacement shown in Figure 2a was measured along the growth direction of the sample, where growth facets could be clearly observed on one side of the crystal (Supplementary Note 1). While it is reasonable to interpret the symmetry breaking by inhomogeneous distribution of charged ionic defects or preferential orientation of MA$^{+1}$ along the growth direction, the inspection of another crystal along a direction orthogonal to the growth direction, Figure 2b, also reveals piezoelectric effect, indicating a complex distribution of defects along different crystal axes. Applying alternating field for a prolonged time homogenises defects distribution throughout the crystal leading to nearly pure electrostrictive effect, Figure 2c.

Another evidence of macroscopic asymmetry of the samples is seen in the voltage signal generated by UV radiation (Supplementary Note 7, Suppl. Fig. 9), and asymmetrical current measured under $E_{ac}$ (Supplementary Note 7, Suppl. Fig. 10). It is noted that asymmetry is expected in samples with different electrodes, as commonly used in optoelectronic applications such as solar cells[30,31,32] However, in our case the two electrodes are the same (both either Pt or C) and the electrode-crystal interface effects[32] can be excluded as a significant contributor to the asymmetry. We emphasize that presented data imply broken macroscopic symmetry of the sample and do not suggest noncentrosymmetric average crystal structure for X=Cl and Br. This can happen if distribution of charged defects is inhomogeneous, as shown next.

**Poling of crystals**

To confirm the likely role of defects distribution in the asymmetry of $MAPbX_3$ samples, we utilized the putative high mobility of ions to manipulate defect distribution by applying $E_{dc}$ for 10 minutes and removing it before applying only $E_{ac}$. By this "poling" procedure[33] we are able to rearrange defects and deterministically induce strong piezoelectric effect and invert its sign, as shown in Figure 2d-e. The evolution in the amplitude of the piezoelectric response immediately after $E_{dc}$ removal indicates that a fraction of defects population displaced by $E_{dc}$ tends to redistribute to a new equilibrium position without external $E_{dc}$ (Supplementary Note 8). Interestingly, a negative poling field $E_{dc}$ (Figure 2e) leads to an increase in the piezoelectric response with time after removal of $E_{dc}$. This could be related to injection of electrons, which interferes with mobility and distribution of ionic defects.

**Frequency dependence of properties**

Analysis of the temporal characteristics of the electro-mechanical response gives further insight into processes involved in the symmetry breaking and the large electro-mechanical response. Data in Figure 1 show time delay between the driving field and strain on the order of several seconds. Hysteresis in electric current-voltage relationship has been reported by many authors for $MAPbX_3$ samples[34] and has been attributed to motion of ionic defects and reorientation of groups of ions. We report a similar hysteresis in the electrically triggered mechanical displacement. Evolution of the displacement hysteresis with the field amplitude is illustrated in Supplementary Note 9. The time lag translates into a large frequency dependence of the piezoelectric, Figure 3a, and electrostrictive responses (Supplementary Note 10, Supplementary Fig. 13a). The broad, non-Debye-like relaxation seen in the permittivity (Supplementary Note 10, Supplementary Fig. 13b) is typical for disordered systems.[35] It suggests different strain and polarization environments for ionic defects throughout the crystal and trapping-detrapping events with different activation energies indicating multiple interactions among defects.[36] The decrease of the mechanical displacement with increasing frequency shows that the electric-field induced strain is dominated by slow defects; the electrical signature of this activity is the current which roughly follows the same trend as the piezoelectric coefficients, Figure 3b. Because of the slowness of the process and long settling times each experiment captures just a snapshot of the sample behaviour. If measurements are continued for sufficiently long time, properties evolve, sometimes radically changing behaviour and even switching internal polarization (Supplementary Note 11). This may explain why so many studies of organometallic halides report mutually inconsistent behaviours. In our experience, this is caused by the strong sensitivity of samples on time and mechanical, electrical, and optical boundary conditions.

**Origin of asymmetry**

We next comment on the nature of species responsible for the strong and slow piezoelectric response. We consider two main candidates: (i) field-driven ionic diffusion via ionic and molecular[31,37] vacancies and (ii) reorientation of organic molecules and defect clusters. There is a considerable controversy about mobility of various defects in $MAPbX_3$.[8] Evidence has been presented about substantial migration of $MA^{+1}$ [31,37] whereas activation energy for $Pb^{+2}$ motion via its vacancies is much higher and motion probably insignificant.[38,6] Reorientation of $MA^{+1}$ is considered (much) faster than the time scales in our experiments.[39] However, because of a presumed high concentration of defects their interaction cannot be neglected and will likely modify characteristic times of individual defects.

It is tempting to correlate the general trend of the piezoelectric coefficients d, Figure 3a, where $d(MAPbCl_3) > d(MAPbBr_3) > d(MAPbI_3)$, with the ionic radius of the halides ($r(Cl)\approx1.81$ Å $< r(Br)\approx 1.96$ Å $< r(I)\approx 2.20$ Å).[11,38] However, this empirical inverse correlation of d with ionic radius should be taken with caution: it may imply that motion or reorientation of $MA^{+1}$ may be easier if the halide ions (or their vacancies) sitting on vertices of the $X_6$ octahedra are smaller. Furthermore, an origin of the spontaneous poling effect in as-grown crystals could be related to orientation of $MA^{+1}$ or defect clusters (e.g., halide vacancy-lead vacancy pair[5]). Despite being fast, reorientation of $MA^{+1}$ molecules under $E_{dc}$ and $E_{ac}$ field probably depends on each molecule's immediate surrounding consisting of slow ionic defects so that the whole charge migration and strain propagation process – diffusion of ions, electronic conductivity which may screen ionic charges and reorientation of organic molecules and defect clusters – are intermingled. How these processes control ion migration is still open to debate.[40] Note that the long times observed in our experiments may suggest migration of ions over longer distances, not only vibration, orientation or jumps within a unit cell.

**Effect of UV light on strain**

The UV light is applied (see Supplementary Note 2) while the sample was driven by electric field, as shown in Figure 4a for $MAPbBr_3$. The data demonstrate that the UV light plays multiple roles in the photo-electro-mechanical coupling. The jump in the displacement signal when the UV light is turned on and off, Figure 4b-4d is caused by the photostrictive effect, already reported for $MAPbX_3$.[41,42] Notably, the data reveal that the UV light polarizes samples. In the sample with graphite electrodes, Figure 4b, pure electrostrictive response observed without UV light becomes partly piezoelectric under the light. Since the electric field is applied without a bias, this is possible only if the sample became polarized by the light (compare with Supplementary Note 7, Supplementary Fig. 9). An indication of the possible mechanism which governs the piezoelectric effect is seen in the sample with Pt electrodes, Figure 4c. Apparent absence of the displacement signal for this sample under $E_{ac}$ field means that the signal is either absent or too small to be detected by the instrument. However, turning on the light reveals both electrostriction and piezoelectricity. The same sample under $E_{ac}+E_{dc}$, Figure 4d, shows that the light increases the piezoelectric signal by factor of four. It has been demonstrated that UV light promotes migration of $MA^{+1}$ ions, but does not affect significantly the motion of $X^{-1}$ ions[31,37]. This selective action of light on ions suggests an important role of $MA^{+1}$ ions in the generation of strain. As the UV light also creates electron-hole pairs the whole charge balance in the crystal will be affected under the light, a new charge environment will develop over time, making deconvolution of different contributions to strain difficult.

**Thermal expansion**

$MAPbX_3$ are known to exhibit a large thermal expansion and a low thermal conductivity[43]. The concern is that the electric-field-generated current could raise the temperature of the sample through Joule heating, with resulting thermal strain interfering with electro-mechanical strain[44] (see Supplementary Note 12.). To gain further insights into the structural mechanisms of the electric-field-induced strain, we investigated $MAPbBr_3$ using single crystal x-ray diffraction[45] (see Methods and Supplementary Note 13).

All the measured Bragg peaks are sharp and single (FWHM ~ 0.1°), indicating a mono-domain nature of $MAPbBr_3$ crystals. Figure 5a shows the time-dependent rocking curve of 004 reflection in the negatively-biased sample and the strain calculated from the shift of the reciprocal space position of 004 peak, Figure 5b. The electric field shifts the position of the rocking curve while preserving its sharp profile, suggesting that the electric-field induced strain is associated with the change of the corresponding lattice parameters. The process responsible for the strain produces therefore an uniform long-range strain field extending over entire crystal, as would be expected for chemical expansion effects[46] and other processes such as local, reversible decomposition-recomposition events under electric field[47, 48]. Figure 5c shows the strain-voltage "butterfly loops", calculated from the reciprocal space position of 004 peak. The agreement with the macroscopic strain measured by LVDT is remarkable.

Next, we investigated if a significant contribution to the strain originates from the thermal expansion of the sample generated by electric-field-induced Joule heating. The pure thermal behaviour of the crystal was probed by measuring the 004 Bragg peak between 20°C and 80°C (Supplementary Note 13, Supplementary Figure 18a). The corresponding thermal expansion coefficient $\alpha = 0.36 \times 10^{-4} \, K$ matches well the literature value.[43] Additionally, we calculated the evolution of integrated intensity of the Bragg peaks form which we can determine the dependence of the integrated intensity on thermal strain. We use the decay of intensity with strain as a "fingerprint" of the thermal origin of the strain. (Supplementary Note 13, Supplementary Figure 18b)

Figure 5c, shows the dependence of relative Bragg peak's intensity on the electric-field induced strain. Notably, the electric-field induced strain is accompanied by the increase of the integrated peak intensity. This behaviour is opposite to the one for thermally induced strain (Supplementary Note 13, Supplementary Figure 18c) which is also shown by the dashed lines in Figure 5d). This indicates that the electric-field-induced strain is different from the strain of purely thermal origin.

We have finally measured sample temperature while applying electric field and compared strains from thermal expansion and electric-field, Figure 5e. The electric-field induced strain is nearly four times larger than the thermal strain.

**Discussion**
Experimental and theoretical investigations of organometallic lead halides mostly focus on electrical properties and electric interactions of ionic defects whereas the strain is usually considered only as an external parameter[23,24,49]. We show that the strain associated with charged ionic defects (effect known as "chemical expansion or contraction"[46]) leads to remarkable electro-mechanical activity resulting in huge macroscopic piezoelectric and electrostrictive responses, much higher than those in oxide perovskites. Local strain resulting from rearrangement of ionic defects under external and internal electric field is expected to interact with and affect recombination of electrons and holes, thus directly influencing performance of optoelectronic devices. A better theoretical understanding of electro-mechanical interactions among defects may help shed light on importance of each defect in the overall performance of these materials. The effects of long-range ionic displacement on electro-mechanical properties revealed in our work should not be limited to organometallic perovskites but are expected to be present in other materials with substantial ionic migration, such as family of fluorites. These effects can be classified under general category of ferroionic-like[50] phenomena, even if ferroelectricity is absent.

## Methods

### Preparation of crystals

MAPbBr$_3$ single crystals were grown using the inverse temperature crystallization method[26] from its saturated solution in dimethylformamide (DMF). MAPbI$_3$ and MAPbCl$_3$ were grown using the identical method, with γ-butyrolactone (GBL) and 50 v/v% DMF / 50 v/v% dimethyl sulfoxide (DMSO). Crystals (Figure 1a, Supplementary Note 1), typically a few millimetres in size, were used as-grown. Electrodes are either sprayed graphite or sputtered platinum, as specified in the text. The electrodes were applied on a parallel pair of {100} faces for X=Br and Cl and parallel to [100] direction for the iodide. Electroded faces for X=Cl and Br are thus either perpendicular or parallel to the crystal growth direction, as shown in Supplementary Note 1.

### Mechanical displacement measurements

Mechanical displacement of the samples was measured using a linear variable differential transformer (LVDT). The LVDT consists of one central primary and two outer secondary coils placed in a tube, forming a transformer. Application of ac voltage on the primary coil induces magnetic field in the transformer. A ferromagnetic core, connected to the sample, moves along the tube axis as the sample size changes under electric field. The magnetic flux changes as the core moves and the voltage difference of the two secondary coils is proportional to the core (sample) displacement. The small voltage difference is read with a lock-in amplifier. The sample is held in contact with the core via a nonmagnetic rod which is supported by two leaf springs which provide restoring force and keep the sample in contact with the core. The displacement on the order of 1 nm can be readily measured, in the frequency range from few mHz to about 10 Hz. (Supplementary Note 2, Supplementary Figure 2; further details can be found in the PhD thesis by Matthew Davis, EPFL; doi: 10.5075/epfl-thesis-3513).

### Current measurements

I-V curves show the relationship between the current I flowing through the sample while a voltage V is applied. The voltage source (Stanford Research DS360) in combination with a voltage amplifier (Trek Model No. 609c6) was used to apply ac voltage to the samples. The current through the samples is estimated by connecting a resistor in series with the sample and measuring voltage drop across the resistor. The oscilloscope (Tektronix, MDO3014) was used for displaying and recording the voltage signals. For the results presented here typical values of the resistor were in the range from 100 to 10000 Ohms. The equivalent circuit and measurement conditions can be found in the PhD thesis of M. Morozov (doi: 10.5075/epfl-thesis-3368). See also Supplementary Note 2.

**XRD measurements**

X-ray diffraction experiments were performed at the custom-built four-circle x-ray diffractometer for in-situ crystallography of functional materials in Tel Aviv University.[45] The diffractometer is equipped with Cu-based microfocus x-ray source, double-crystal monochromator (fully suppressing Kalpha2 line), motorized Eulerian cradle, two-dimensional pixel area detector (PILATUS 1M) and the high-temperature temperature cell (Forvis Technologies). Electric field was produced using HAMEG electric signal arbitrary function generator and amplified using Matsusada high-voltage amplifier. The entire instrument is additionally equipped with various electronic circuits for the synchronization between applied electric field, diffractometer control software and the x-ray detector. Further experimental details are available in the Supplementary Note 13. For a schematic presentation see Supplementary Figure 17.

**References:**

1. Chouhan, L., Ghimire, S., Subrahmanyam, C., Miyasaka, T. & Biju, V. Synthesis, optoelectronic properties and applications of halide perovskites. *Chem. Soc. Rev.* **49**, 2869–2885 (2020).

2. Zhang, T., Hu, C. & Yang, S. Ion Migration: A "Double-Edged Sword" for Halide-Perovskite-Based Electronic Devices. *Small Methods* **4**, 1900552 (2020).

3. Zhang, X., Shen, J.-X., Turiansky, M. E. & Van de Walle, C. G. Minimizing hydrogen vacancies to enable highly efficient hybrid perovskites. *Nat. Mater.* **20**, 971–976 (2021).

4. Zhou, Y., Poli, I., Meggiolaro, D., De Angelis, F. & Petrozza, A. Defect activity in metal halide perovskites with wide and narrow bandgap. *Nat Rev Mater* **6**, 986–1002 (2021).

5. Keeble, D. J. *et al.* Identification of lead vacancy defects in lead halide perovskites. *Nat Commun* **12**, 5566 (2021).

6. Eames, C. *et al.* Ionic transport in hybrid lead iodide perovskite solar cells. *Nature communications* **6**, 7497 (2015).

7. Lee, J.-W., Kim, S.-G., Yang, J.-M., Yang, Y. & Park, N.-G. Verification and mitigation of ion migration in perovskite solar cells. *APL Materials* **7**, 041111 (2019).

8. Mosconi, E. & De Angelis, F. Mobile Ions in Organohalide Perovskites: Interplay of Electronic Structure and Dynamics. *ACS Energy Lett.* **1**, 182–188 (2016).

9. Cordero, F., Craciun, F. & Trequattrini, F. Ionic Mobility and Phase Transitions in Perovskite Oxides for Energy Application. *Challenges* **8**, 5 (2017).

10. Kieslich, G., Sun, S. & Cheetham, A. K. Solid-state principles applied to organic–inorganic perovskites: new tricks for an old dog. *Chem. Sci.* **5**, 4712–4715 (2014).

11. Shannon, R. D. Revised Effective Ionic Radii and Systematic Studies of Interatomic Distances in Halides and Chaleogenides. *Acta Cryst. A* **32**, 751 (1976).

12. Nowick, A. S. & Heller, W. R. Dielectric and anelastic relaxation of crystals containing point defects. *Adv. Phys.* **14**, 101–166 (1965).

13. Chen, B. *et al.* Large electrostrictive response in lead halide perovskites. *Nature Materials* **17**, 1020 (2018).

14. Shu, L. *et al.* Photoflexoelectric effect in halide perovskites. *Nat. Mater.* 1–5 (2020) doi:10.1038/s41563-020-0659-y.

15. Newnham, R. E. *Properties of Materials: Anisotropy, Symmetry, Structure*. (Oxford University, 2005).

16. Onoda-Yamamuro, N., Matsuo, T. & Suga, H. Dielectric study of CH3NH3PbX3 (X = Cl, Br, I). *Journal of Physics and Chemistry of Solids* **53**, 935–939 (1992).

17. Coll, M. *et al.* Polarization Switching and Light-Enhanced Piezoelectricity in Lead Halide Perovskites. *The journal of physical chemistry letters* **6**, 1408–13 (2015).

18. Rakita, Y. *et al.* Tetragonal CH3NH3PbI3 is ferroelectric. *Proceedings of the National Academy of Sciences of the United States of America* **114**, E5504–E5512 (2017).

19. Kim, Y.-J. *et al.* Piezoelectric properties of CH3NH3PbI3 perovskite thin films and their applications in piezoelectric generators. *J. Mater. Chem. A* **4**, 756–763 (2016).

20. Garten, L. M. *et al.* The existence and impact of persistent ferroelectric domains in MAPbI3. *Science Advances* **5**, eaas9311 (2019).

21. Fan, Z. *et al.* Ferroelectricity of CH3NHPbI3 Perovskite. *J. Phys. Chem. Lett.* **6**, 1155–1161 (2015).

22. Rappe, A. M., Grinberg, I. & Spanier, J. E. Getting a charge out of hybrid perovskites. *Proc. Natl. Acad. Sci. U. S. A.* **114**, 7191–7193 (2017).

23. Zhu, C. *et al.* Strain engineering in perovskite solar cells and its impacts on carrier dynamics. *Nat Commun* **10**, 1–11 (2019).

24. Liu, D. *et al.* Strain analysis and engineering in halide perovskite photovoltaics. *Nat. Mater.* **20**, 1337–1346 (2021).

25. Zhang, Y., Yang, Y. & Wang, Z. L. Piezo-phototronics effect on nano/microwire solar cells. *Energy Environ. Sci.* **5**, 6850–6856 (2012).

26. Saidaminov, M. I. *et al.* High-quality bulk hybrid perovskite single crystals within minutes by inverse temperature crystallization. *Nature communications* **6**, 7586 (2015).

27. Newnham, R. E., Sundar, V., Yimnirun, R., Su, J. & Zhang, Q. M. Electrostriction: Nonlinear Electromechanical Coupling in Solid Dielectrics. *The Journal of Physical Chemistry B* **101**, 10141–10150 (1997).

28. Jang, S. J., Uchino, K., Nomura, S. & Cross, L. E. Electrostrictive behvior of lead magnesium nobate based ceramic dielectrics. *Ferroelectrics* **27**, 31–34 (1980).

29. Park, S.-E. & Shrout, T. R. Ultrahigh Strain and Piezoelectric Behavior in Relaxor Based Ferroelectric Single Crystals. *Journal of Applied Physics* **82**, 1804–1811 (1997).

30. Deng, Y., Xiao, Z. & Huang, J. Light-Induced Self-Poling Effect on Organometal Trihalide Perovskite Solar Cells for Increased Device Efficiency and Stability. *Advanced Energy Materials* **5**, 1500721 (2015).

31. Liu, Y. *et al.* Hysteretic Ion Migration and Remanent Field in Metal Halide Perovskites. *Advanced Science* **7**, 2001176 (2020).

32. Yang, M.-M. *et al.* Piezoelectric and pyroelectric effects induced by interface polar symmetry. *Nature* **584**, 377–381 (2020).

33. Yuan, Y. & Huang, J. Ion Migration in Organometal Trihalide Perovskite and Its Impact on Photovoltaic Efficiency and Stability. *Acc. Chem. Res.* **49**, 286–293 (2016).

34. Tress, W. *et al.* Understanding the rate-dependent J–V hysteresis, slow time component, and aging in CH3NH 3PbI3 perovskite solar cells: the role of a compensated electric field. *Energy Environ. Sci.* **8**, 995–1004 (2015).

35. Jonscher, A. K. *Dielectric relaxation in solids*. (Chelsea Dielectric Press, 1983).

36. Knop, O., Wasylishen, R. E., White, M. A., Cameron, T. S. & Oort, M. J. M. V. Alkylammonium lead halides. Part 2. CH3NH3PbX 3 (X = Cl, Br, I) perovskites: cuboctahedral halide cages with isotropic cation reorientation. *Can. J. Chem.* **68**, 412–422 (1990).

37. Liu, Y. *et al.* Direct Observation of Photoinduced Ion Migration in Lead Halide Perovskites. *Advanced Functional Materials* **31**, 2008777 (2020).

38. Futscher, M. H. *et al.* Quantification of ion migration in CH3NH3PbI3 perovskite solar cells by transient capacitance measurements. *Mater. Horiz.* **6**, 1497–1503 (2019).

39. Cordero, F. Cation reorientation and octahedral tilting in the metal-organic perovskites MAPI and FAPI. *Journal of Alloys and Compounds*.

40. Herz, L. M. Charge-Carrier Mobilities in Metal Halide Perovskites: Fundamental Mechanisms and Limits. *ACS Energy Lett.* **2**, 1539–1548 (2017).

41. Wei, T.-C. *et al.* Photostriction of CH3NH3PbBr3 Perovskite Crystals. *Advanced Materials* **29**, 1701789 (2017).

42. Lv, X. *et al.* Giant Bulk Photostriction and Accurate Photomechanical Actuation in Hybrid Perovskites. *Advanced Optical Materials* **9**, 2100837.

43. Ge, C. *et al.* Ultralow Thermal Conductivity and Ultrahigh Thermal Expansion of Single-Crystal Organic–Inorganic Hybrid Perovskite CH3NH3PbX3 (X = Cl, Br, I). *J. Phys. Chem. C* **122**, 15973–15978 (2018).

44. Tsai, H., Nie, W. & Mohite, A. D. Response to Comment on "Light-induced lattice expansion leads to high-efficiency solar cells". *Science* **368**, (2020).

45. Gorfman, S. *et al.* Multipurpose diffractometer for in situ X-ray crystallography of functional materials. *J Appl Cryst* **54**, 914–923 (2021).

46. Marrocchelli, D., Bishop, S. R., Tuller, H. L. & Yildiz, B. Understanding Chemical Expansion in Non-Stoichiometric Oxides: Ceria and Zirconia Case Studies. *Advanced Functional Materials* **22**, 1958–1965 (2012).

47. Jo, Y.-R., Tersoff, J., Kim, M.-W., Kim, J. & Kim, B.-J. Reversible Decomposition of Single-Crystal Methylammonium Lead Iodide Perovskite Nanorods. *ACS Cent Sci* **6**, 959–968 (2020).

48. Chen, S. *et al.* Atomic scale insights into structure instability and decomposition pathway of methylammonium lead iodide perovskite. *Nat Commun* **9**, 4807 (2018).

49. Xue, D.-J. *et al.* Regulating strain in perovskite thin films through charge-transport layers. *Nat Commun* **11**, 1514 (2020).

50. Liu, Y. *et al.* Ferroic Halide Perovskite Optoelectronics. *Advanced Functional Materials* **31**, 2102793 (2021).

**Data availability**

All data are presented in the text and Supplementary Notes are available from M.V, S.G. and D.D on request.

**Acknowledgements**

M.V. and D.D. acknowledge financial support of the ONR Global (Grants N62909-19-1-2092 and N62909-20-1-2083). The work of M.K., E.H. and L.F. was supported by ERC Advanced Grant "PICOPROP" (Grant No. 670918). S.G. acknowledges financial support of Israel Science Foundation (Grants No 1561/18, 2247/18 and 3455/21). Contribution of Andrea Charrier at early stages of this work is gratefully acknowledged.

**Author contributions**

M.V. performed all electrical, mechanical displacement and optical experiments, analysed and interpreted the data. M.K. and E.H. grew crystals. L.F. coordinated the crystal growth work. D.S. and S.G. performed X-ray diffraction measurements, analysed and interpreted the data. L.R. constructed the setup for UV measurements. All coauthors discussed results. D.D. conceived the idea, supervised the work and participated in the experimental work, data analysis and interpretation. M.V., S.G., and D.D. wrote the paper with input from all coauthors.

**Competing interests**

The authors declare no competing interests.

Figures and figure captions:

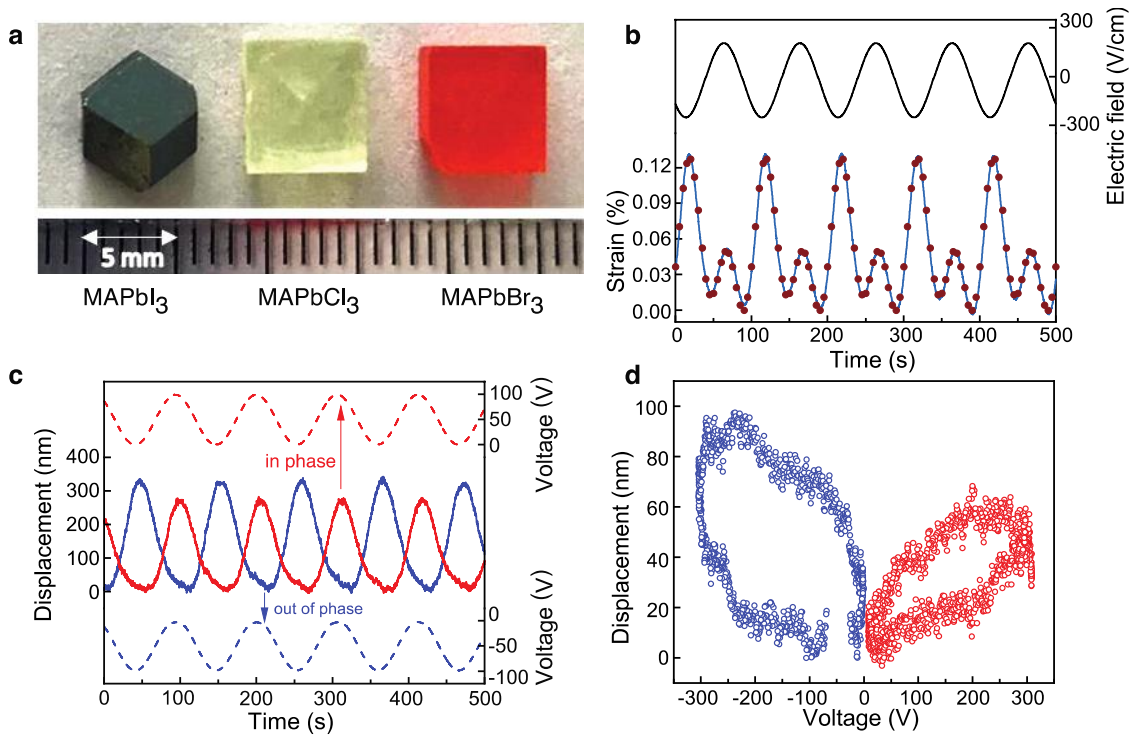

**Figure 1. Three aspects of induced piezoelectricity**. **a,** Examples of MAPbX$_3$ crystals. The sides of the crystals in the plane of the image are perpendicular to the growth direction. The field was applied along the growth direction for X=Cl and Br and perpendicular to it for X=I. **b,** Coexistence of electrostriction and piezoelectricity in an MAPbCl$_3$ crystal at small E$_{dc}$ (–34 V/cm) and large E$_{ac}$ (≈250 V/cm). The field was applied along the thickness t=2.15 mm. **c,** Change in the sign of the bias field E$_{dc}$ flips the phase of the displacement in an MAPbBr$_3$ sample indicating that the piezoelectric coefficient d is controlled by E$_{dc}$. t=1.32 mm. **d,** Asymmetric and hysteretic displacement-voltage relationship in an MAPbI$_3$ crystal. t=5.68 mm. All samples had sprayed graphite electrodes.

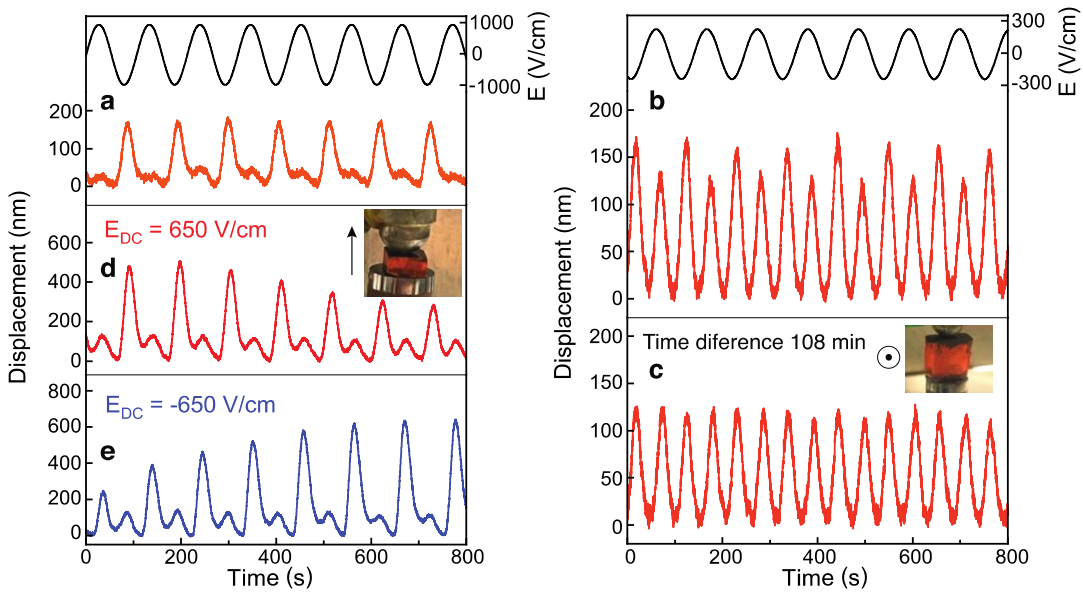

**Figure 2. Asymmetry and poling**. Electromechanical displacement of: **a,** an as received MAPbBr$_3$ sample with only E$_{ac}$ applied along the growth direction. The piezoelectric signal is dominant. **b,** another as received MAPbBr$_3$ sample, measured applying only E$_{ac}$ along a direction orthogonal to the growth direction. The response is dominantly electrostrictive but with a non-negligible piezoelectric contribution. **c,** the same sample and direction as in **(b)** after driving the sample for additional 108 minutes with E$_{ac}$. The asymmetry and piezoelectric response are nearly eliminated by homogenizing the defect distribution resulting in only electrostrictive response. **d,** the sample as in **(a)**, after being poled by +E$_{dc}$ for ten minutes. The measurements were performed by applying only E$_{ac}$. **e,** the same sample as in **(a,d)**, after being poled by –E$_{dc}$ for ten minutes and then driven by applying only E$_{ac}$. Note expected change in the phase between **(d)** and **(e)**. Both MAPbBr$_3$ samples had graphite sprayed electrodes. Thickness t=1.54 mm along the growth direction in **(a)**, **(d)**, **(e)** and V$_{ac}$=150 V. Thickness t=6.40 mm along the direction orthogonal to the growth direction in **(b-c)** and V$_{ac}$ = 150 V. Insets in **Figure 2d** and **2c** show crystals with electrodes deposited on faces perpendicular and parallel, respectively, to the growth direction.

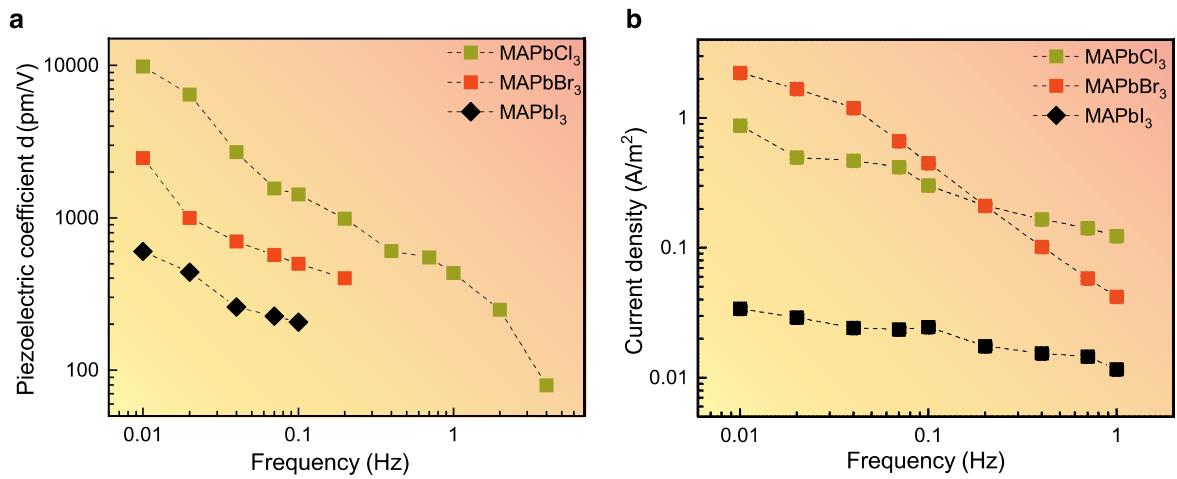

**Figure 3. Frequency dependence of properties. a,** piezoelectric coefficient and **b,** the alternating current in MAPbX$_3$ crystals. All samples had graphite sprayed electrodes. In **(a)** MAPbCl$_3$ thickness t=3.8 mm and V$_{ac}$=150 V applied along the growth direction; MAPbBr$_3$ t=1.87 mm, V$_{ac}$=100 V applied along the growth direction; MAPbI$_3$ t=5.68mm, V$_{ac}$=150 V applied perpendicular to the growth direction. In **(b)** MAPbCl$_3$ t=3.87 mm, MAPbBr$_3$ t=4.61 m, MAPbI$_3$, t=5.68 mm. In all three cases V$_{ac}$=15 V, applied perpendicular to the growth direction.
19

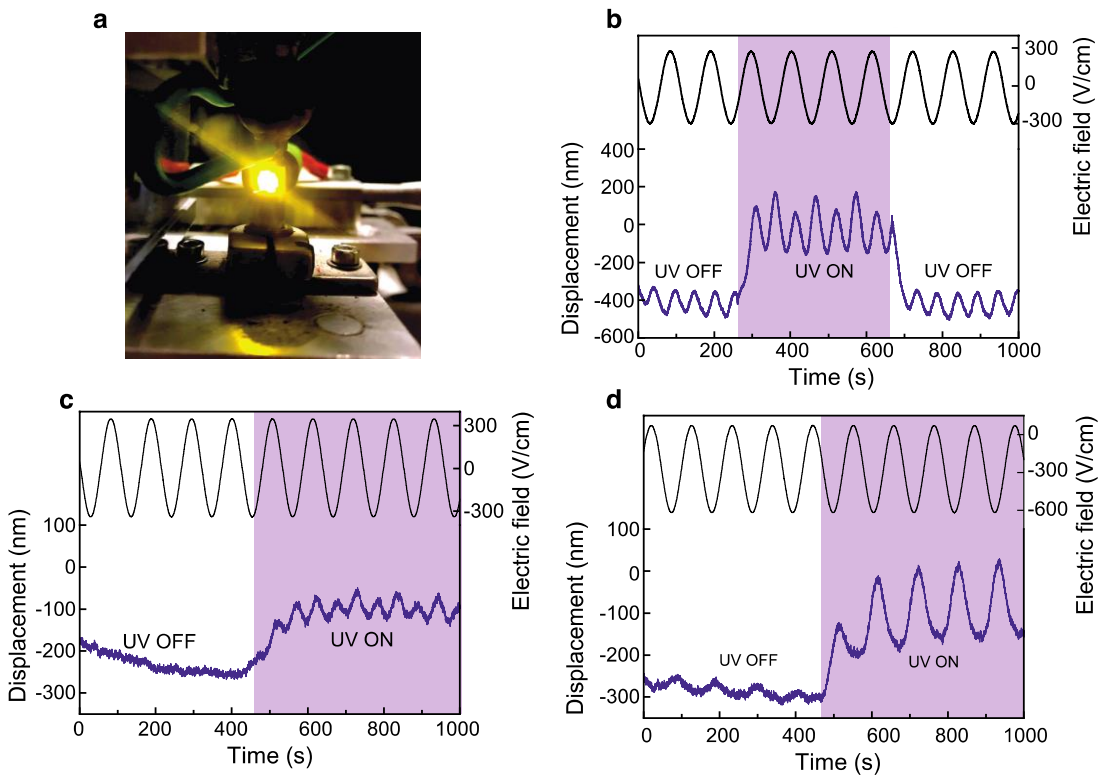

**Figure 4. Effect of UV light on strain in MAPbBr$_3$. a**, a crystal in the LVDT setup with the fibre optic UV light guide in the back. The light beam is perpendicular to the plane of the picture. **b,** electro-mechanical displacement measured in a sample with sprayed graphite electrodes, thickness t=5.23 mm and V$_{ac}$ =150 V applied perpendicular to the growth direction. **c,** displacement in a sample with sputtered Pt electrodes, t= 3.59 mm and V$_{pp}$=125 V applied along the growth direction. **d,** the same as in (**c**) with additional V$_{dc}$= –100 V.

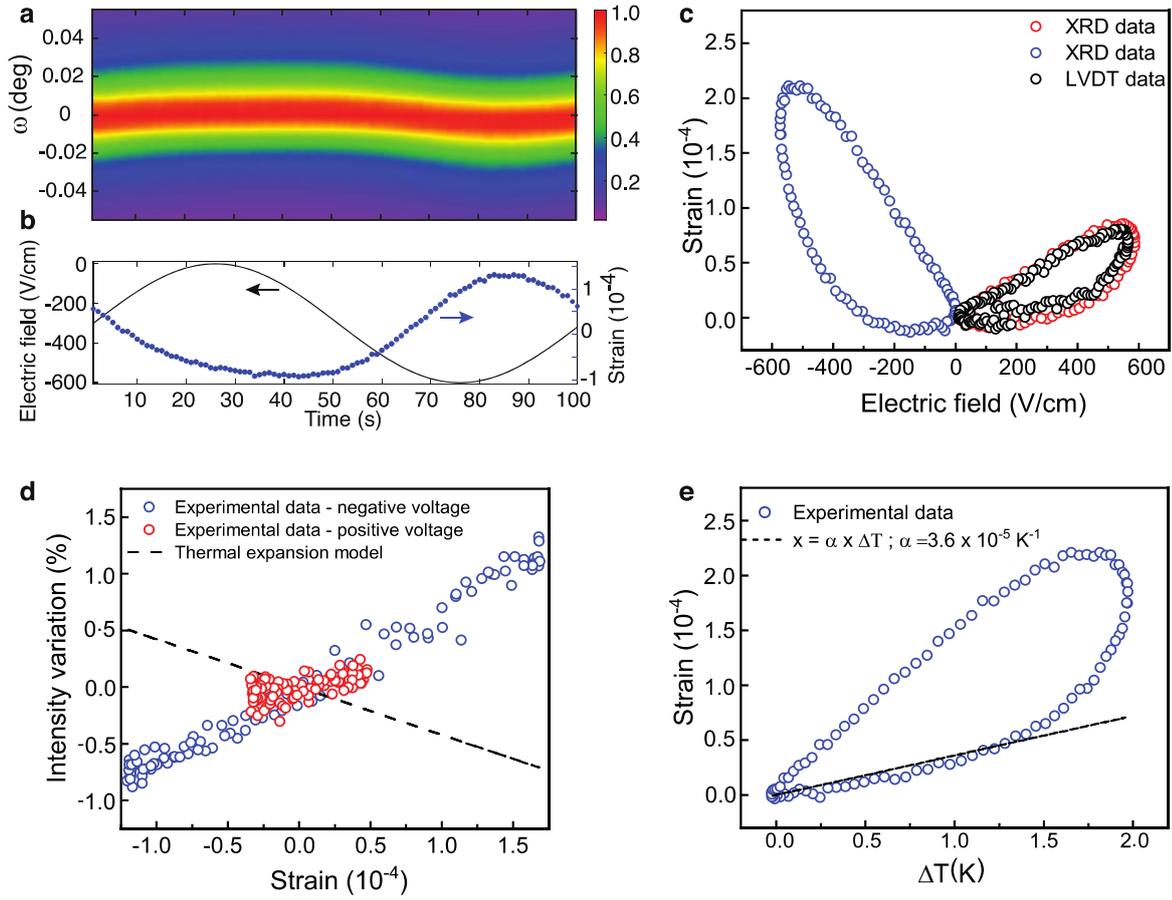

**Figure 5. Thermal expansion effects in MAPbBr$_3$. a,** False colour map of the normalized x-ray diffraction intensity of rocking curve of 004 Bragg reflection. ω is the rocking angle. **b,** time-dependence of the applied voltage and resulting strain. Note that the zero point of strain is arbitrarily chosen. **c,** Voltage dependence of the strain, extracted from the displacement of the rocking curve of 004 Bragg reflection under applied electric field. Red dots correspond to the positive offset voltage, the blue dots to the negative offset voltage (the same data as in (**b**)), and the black dots correspond to experimental data obtained by LVDT. **d,** The dependence of integrated intensity of 004 reflection on the strain during the application of external electric field. Red dots represent the case of positive offset voltage $V_{dc}= +50$ V, blue dots the case of negative offset voltage $V_{dc}= -50$ V, the dashed line describes the strain expected from the thermal expansion model. **e,** The dependence of strain on the variation of temperature at the sample. The dashed line depicts the dependence of strain on the temperature expected from thermal expansion. The sample had sprayed graphite electrodes and thickness t=1.74 mm.

# Supplementary Information

## "Forbidden" polarisation and extraordinary piezoelectric effect in organometallic lead halide perovskites

Milica Vasiljevic[1], Márton Kollár [2,4], David Spirito[3], Lukas M. Riemer[1], László Forró [2,5], Endre Horváth [2,4,6], Semën Gorfman[3], Dragan Damjanovic[1,*]

1. Group for Ferroelectrics and Functional Oxides, Ecole Polytechnique Fédérale de Lausanne 1015 Lausanne, Switzerland
2. Laboratory of Physics of Complex Matter, Ecole Polytechnique Fédérale de Lausanne, 1015 Lausanne, Switzerland
3. The Department of Materials Science and Engineering, Tel Aviv University, Ramat Aviv, Tel Aviv 6997801, Israel
4. KEP Technologies, 1228 Plan-les-Ouates, Switzerland
5. Stavropoulos Center for Complex Quantum Matter, University of Notre Dame, IN 46556, USA
6. Laboratory for Quantum Magnetism, Institute of Physics, École Polytechnique Fédérale de Lausanne, CH1015 Lausanne, Switzerland

*Email: dragan.damjanovic@epfl.ch

List of Supplementary Notes

1. Crystal morphology
2. Mechanical displacement and current measurements and UV illumination of the samples.
3. Calculation of electrostrictive and piezoelectric coefficients
4. Electrostrictive coefficients for various materials versus permittivity
5. Very large values of d in $MAPbBr_3$
6. Discrepancy between experimental d and calculated from $2ME_{dc}$
7. Evidence of macroscopic asymmetry.
8. Decay of piezoelectric response after poling
9. Evolution of hysteresis with electric field
10. Frequency dependence of the response
11. Change of internal polarization and properties with time
12. Asymmetrical currents, Joule heating and thermal expansion
13. x-ray diffraction

**Supplementary Note 1. Crystal morphology**

The growth habits of MaPbX$_3$ crystals for X=Cl and Br are the same, reflecting their common cubic *Pm*-3*m* symmetry, as illustrated in Figure 1a. The growth habit of iodide is more complex (Figure 1a) and has not been studied in details here. Identification of crystallographic faces for MAPbI$_3$ can be found in supplementary ref. [1].

Supplementary Figure 1 illustrates crystal facets formed on one of {100} cubic planes of the crystal. We refer to the direction perpendicular to this plane as the "crystal growth direction". This face of the crystal serves as a convenient reference for orientation of the crystal. The facets form on the seeded side of the crystal, oriented toward the bottom of the vessel during the growth.

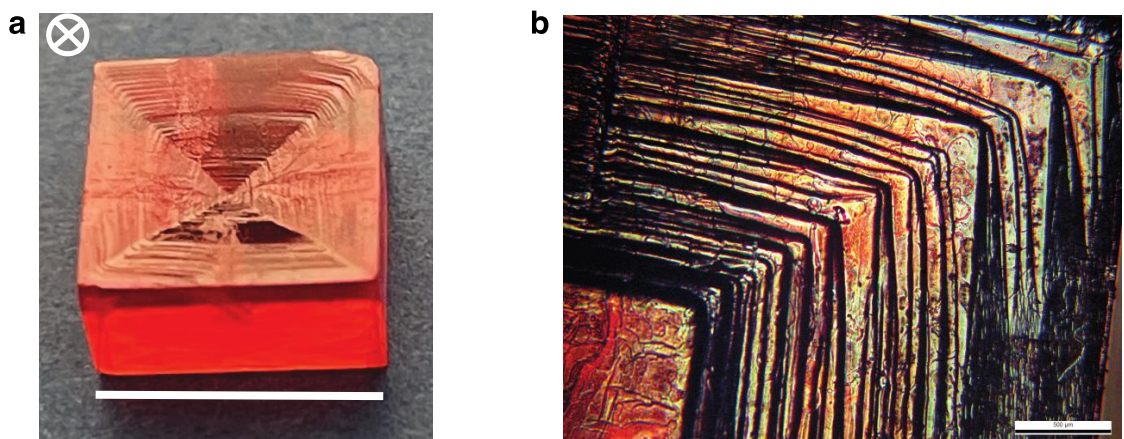

**Supplementary Figure 1. MAPbBr$_3$ crystal. a,** Illustration of the crystal habit and the crystal growth direction indicated by the tail of arrow. The crystal was seeded on this side. Six orthogonal sides of the cuboid belong to the family of {100} planes of the cubic symmetry. The (100) "terraces" are visible on the "growth side." The scale bar is 7 mm. **b,** close-up view of one corner of the crystal on the growth side. The scale bar is 500 µm. The most displacement measurements were made along the growth direction, while the current was usually measured along one of the perpendicular <100> directions.

**Supplementary Note 2. Mechanical displacement and current measurements and UV illumination of the samples.**

*Mechanical displacement measurements*. See also Methods. LVDT used in this work was model "050 HR" from Schaevitz. It has a sensitivity of around 228 mV/V/mm with a small deviation from linearity ( ±0.1%) in a linear range of 0.6 mm. Displacements of 1nm could be measured with a help of a lock-in amplifier.[2] The output voltage-to-displacement conversion factor was 1.55x10$^{-6}$ m/V. The LVDT set-up with a sample is shown in Supplementary Figure 2a. Supplementary Figure 2b shows the closeup of the LVDT setup with the head of fibre optic UV light guide near the sample, as used in some experiments. Note that we measure longitudinal electro-mechanical effect where field is applied and displacement measured along the same direction.

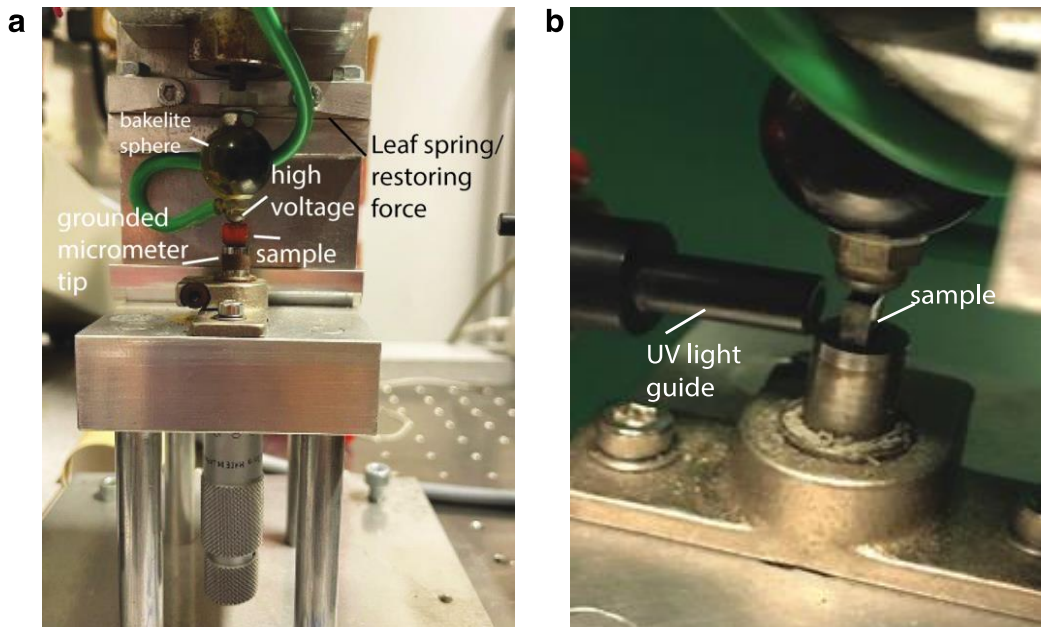

**Supplementary Figure 2**. **a,** the LVDT setup with an MAPbBr$_3$ sample. See also Methods. **b,** a closeup of the LVDT setup with the head of the fibre optic UV light guide near an MAPbCl$_3$ sample.

*Illumination with UV light*. Samples were illuminated with UV light using a Hg Arc Lamp UV Source (Oriel Instruments; Power Supply Model 68806; a Series Q Lamp Housing Model 60000; lamp model 6281). The source power is 100 W. The head of the fibre guiding the light from the UV source is positioned to illuminate the sample from a close distance (few millimetres), as shown in Supplementary Figure 2b. The lamp shutter is controlled in on-off mode using a relay driven by square-wave unipolar voltage (V=10 V) with variable frequency (10 mHz in this work) generated by a frequency generator (Agilent 33120A).

*I-V current measurements.* See also Methods. The current through the samples was measured by adding a resistor R in series with the sample with capacitance C, Supplementary Figure 3. The resistor serves only to measure the current determined by the voltage drop across the sample (capacitor.) This condition is satisfied if R<<1/$\omega$C, where $\omega$ is the angular frequency of the voltage signal.[3] Taking this condition into account, the size of the resistor has been adapted to the sensitivity of the instrument (V$_{out}$=RI).

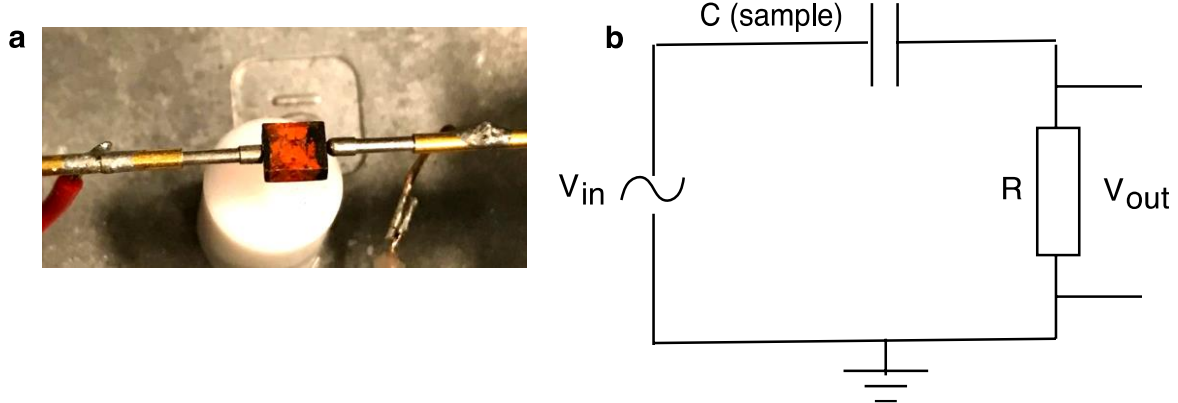

**Supplementary Figure 3**. **a,** an MAPbBr₃ sample in the sample holder for I-V measurements. **b,** equivalent circuit for I-V measurements.

**Supplementary Note 3. Calculation of electrostrictive and piezoelectric coefficients**

A general equation for strain induced by electric field $E_{ac}(t) = E_{dc} + E_o \sin(\omega t)$ is given by (up to second order):

$$x(t) = dE_{ac} + ME_{ac}^2 \qquad (1)$$

or

$$x(t) = dE_o\sin(\omega t) + dE_{dc} + ME_{dc}^2 + 2E_{dc}ME_o\sin(\omega t) + M(E_o\sin(\omega t))^2 = (dE_{dc} + ME_{dc}^2) + (d + 2E_{dc}M)E_o\sin(\omega t) + M(E_o\sin(\omega t))^2 \qquad (2)$$

where d is the piezoelectric coefficient (if the material is noncentrosymmetric) and M is the electrostriction coefficient. The first term on the right, $(dE_{dc} + ME_{dc}^2)$, in (2) describes the total static strain. The second, linear, term on the right is the total piezoelectric-like response, where $(2E_{dc}M)E_o\sin(\omega t)$ is the biased electrostriction and $dE_o\sin(\omega t)$ is the proper piezoelectric strain. The last, quadratic term, is pure electrostriction. An illustration of the total strain is given in Supplementary Figure 4.

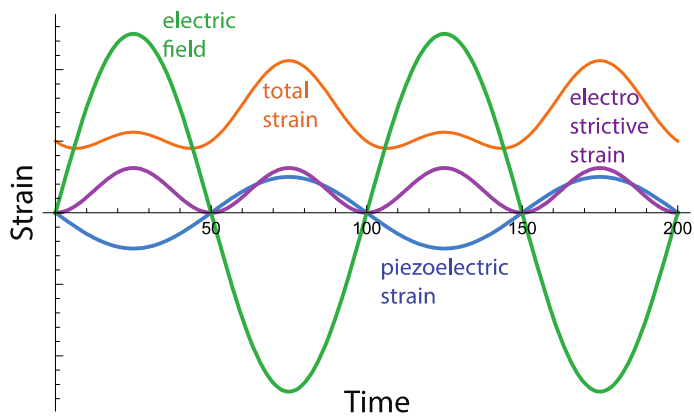

**Supplementary Figure 4**. Illustration of different terms from equation (1), for $E_{dc}=0$ and $d<0$.

To determine d and M from measured strain x(t), one needs two values of strain, x(t1) and x(t2). The solution is given by:

$$M = \frac{x[t1]}{E_0(E_{dc}+E_0\text{Sin}[\omega t1])(\text{Sin}[\omega t1]-\text{Sin}[\omega t2])} - \frac{x[t2]}{E_0(E_{dc}+E_0\text{Sin}[\omega t2])(\text{Sin}[\omega t1]-\text{Sin}[\omega t2])} \quad (3)$$

$$d = \frac{(E_{dc}+E_0\text{Sin}[\omega t1])x[t2]}{E_0(\text{Sin}[\omega t1]-\text{Sin}[\omega t2])(E_{dc}+E_0\text{Sin}[\omega t2])} - \frac{(E_{dc}+E_0\text{Sin}[\omega t2])x[t1]}{E_0(E_{dc}+E_0\text{Sin}[\omega t1])(\text{Sin}[\omega t1]-\text{Sin}[\omega t2])} \quad (4)$$

It is convenient to take t1 and t2 that correspond to local maxima of x(t) which appear at $\omega t1=\pi/2$ and $\omega t2=3\pi/2$. Then (3) and (4) become (for $E_0 \neq E_{dc}$)

$$M = \frac{x[1/4f]}{2E_0(E_{dc}+E_0)} - \frac{x[3/4f]}{2E_0(E_{dc}-E_0)} \quad (5)$$

$$d = \frac{(E_{dc}+E_0)x[3/4f]}{2E_0(E_{dc}-E_0)} - \frac{(E_{dc}-E_0)x[1/4f]}{2E_0(E_{dc}+E_0)} \quad (6)$$

where f is the driving frequency ($\omega t1=2\pi ft1=\pi/2$; $\omega t2=2\pi ft2=3\pi/2$). In the special case when $E_{dc}=E_o$, the points $\omega t1=\pi/2$ and $\omega t2=3\pi/2$ lead to only one equation:

$$2dE_o + 4ME_o^2 = x[1/4f] \quad (7)$$

because $x[3/4f] = 0$, see Supplementary Figure 5. Thus, to determine M and d it is necessary to take another time point and use equations (3) and (4). If d=0, then $4ME_o^2 = x[1/4f]$

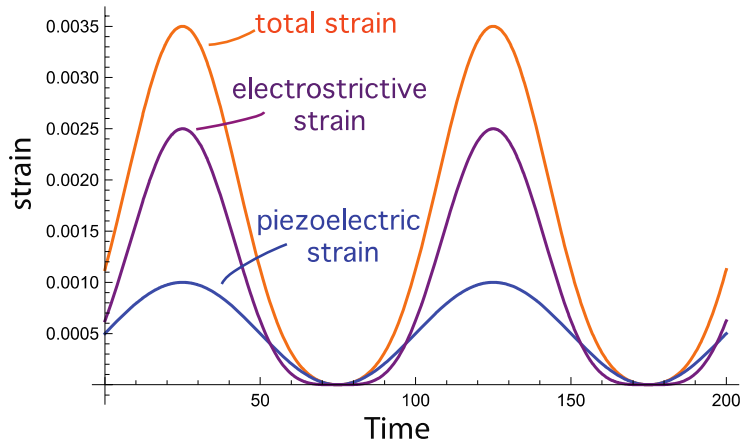

**Supplementary Figure 5**. Illustration of different terms from equation (1) when $E_{dc}=E_o$

If there is a phase angle δ between the electric field and strain (in other words, if d and M are complex), then taking (5) and (6) and assuming that maxima are at $\omega t=\pi/2, 3\pi/2$, will lead to errors. These can be estimated if δ is known:
For example, $dE_o\sin(\omega t)$ will have a maximum at $\omega t + \delta$ so that $\sin(\omega t + \delta) = \cos(\delta)$ for $\omega t = \pi/2$ and $\sin(\omega t + \delta) = -\cos(\delta)$ for $\omega t = 3\pi/2$. Thus, $E_o$ in equations (5) and (6) should be replaced by $E_o\cos(\delta)$. For a phase angle of $\pi/7$ and $\pi/11$ in Fig. 1a, for example, $\cos(\delta)$ is, respectively, 0.9 and 0.95, that is, the error in the associated terms in equations (5) and (6) is 10% and 5%, As we are concerned here with the order of magnitude of coefficients this correction was not made.

## Supplementary Note 4. Electrostrictive coefficients for various materials versus permittivity

Electrostrictive coefficient M is plotted as a function of relative permittivity $\varepsilon$ for a range of materials in Supplementary Figure 6. The weak-field relative permittivity of MAPbX$_3$ measured at 1 kHz[4] is ~50 and the values displayed here (red lines) were measured at 0.01 Hz at relatively large voltages (see text and Ref.[5]) where values of permittivity are not readily available. The motivation to plot M versus permittivity is driven by fundamental relation between the field electrostriction coefficient M and polarization electrostriction coefficient Q[6]:

$$x = QP^2 = ME^2 \text{ and } P \approx \varepsilon E \Rightarrow M = \varepsilon Q$$

where P is the polarization (induced or spontaneous). The relationship suggests that most of M's dependence on frequency and field should arise from the permittivity. That is, in contrast to the Q coefficient, M is controlled by extrinsic contributions.[6] The data correlation in Supplementary Figure 6 suggests that the apparent relative permittivity of MAPbX$_3$ at low frequencies should be on the order of $10^4$-$10^6$ (because of high ionic contributions) and this is seen by direct measurements of the permittivity (shown below in Supplementary Note 10).

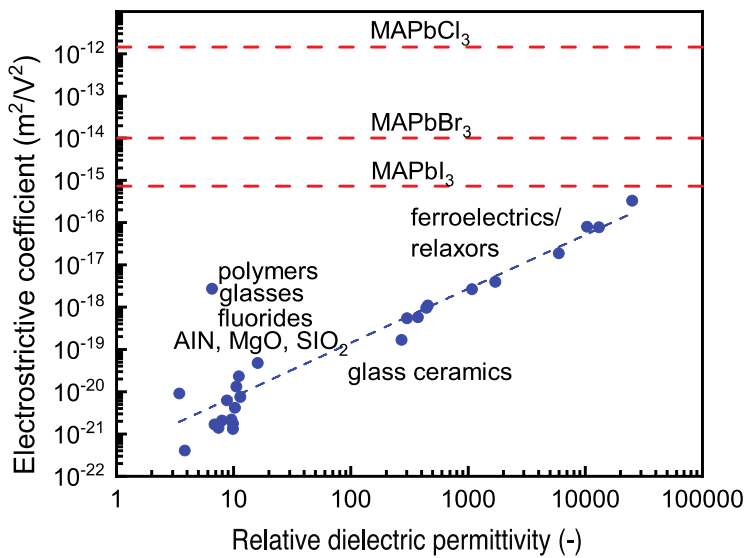

**Supplementary Figure 6**. Electrostrictive coefficient M in function of relative dielectric permittivity for selected materials[6]. The values for MaPbI$_3$ is from Ref.[5] and values for MaPbBr$_3$ and MAPbCl$_3$ are from the present work. Compare with Supplementary Figure 13.

## Supplementary Note 5. Very large values of the effective longitudinal piezoelectric coefficient (d) in MAPbBr$_3$

The values of the electrostrictive and piezoelectric coefficients for a given MAPbX$_3$ composition depend on sample, its history, frequency, amplitude of the driving field and other parameters. We have observed variations in the values of d coefficient in MAPbBr$_3$ by more than factor of two. Compare, for example, data in Figure 5c (where d for the negative field is on the order of 4000 pm/V) and data in Supplementary Figure 7.

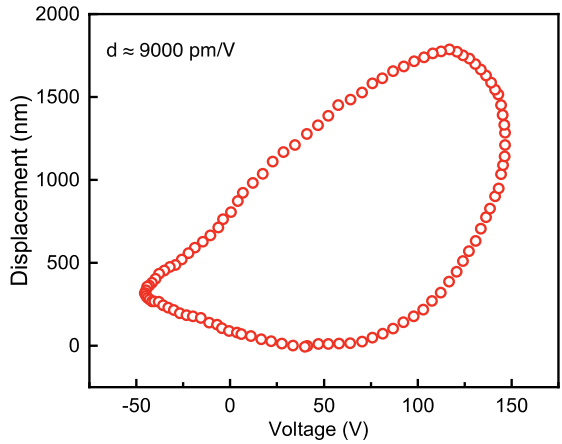

**Supplementary Figure 7.** Displacement versus voltage measured at 10 mHz in an MAPbBr$_3$ sample. The sample had sprayed graphite electrodes, thickness t=2.76 mm, and the driving voltage was $V_{ac}$=100 V, $V_{dc}$= 50V. The effective piezoelectric coefficient calculated form the slope of the linear part of displacement is over 9000 pm/V.

**Supplementary Note 6. Comment on discrepancy between experimental d and calculated from 2ME$_{dc}$**

The piezoelectric coefficient d was determined from experimental data using relationship $x = ME_{ac}^2 + dE_{ac}$ (see Supplementary Note 3) and was compared with values expected from d= $2ME_{dc}$ as a function of bias voltage $V_{dc}$. We find discrepancy which increases irregularly with the strength of $V_{dc}$, Supplementary Figure 8.

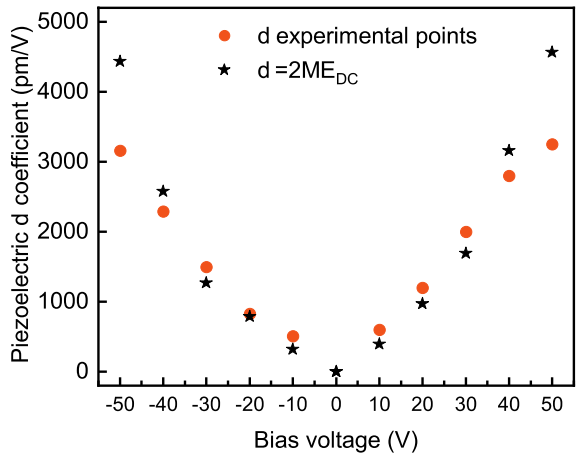

**Supplementary Figure 8.** Comparison of the values of d coefficient determined from $x = ME_{ac}^2 + dE_{ac}$ with values expected from d=2ME$_{dc}$ for an MAPbBr$_3$ sample with thickness t=1.32 mm. The driving voltage $V_{ac}$ = 50 V. For $V_{dc}$= ±50V, the expected value d ~5000 pm/V is 1.5 times higher than the value obtained from the experimental data, while for the lower $V_{dc}$ experimental data are higher than estimated values. This discrepancy suggests that the sample has an internal bias which evolves with application of external field. The sample had sprayed graphite electrodes.

**Supplementary Note 7. Evidence of macroscopic asymetry of the samples**

When an MAPbX$_3$ sample is illuminated with UV light (Supplementary Figure 9), electric potential difference is generated between the electrodes even when electrodes are made of the same material. In experiments performed here two opposites sides perpendicular to the crystal growth direction were sprayed with graphite. When the sample is flipped over, the potential difference changes sign and amplitude. The potential difference suggests that there is a built-in asymmetry in the samples, possibly caused by inhomogeneous distribution of defects.

Another indication of built-in asymmetry in the samples is electric current asymmetry, Supplementary Figure 10. The zero-current voltage offset should also be noted, indicating internal bias, possibly caused by illumination of the sample by visible light during the measurements.

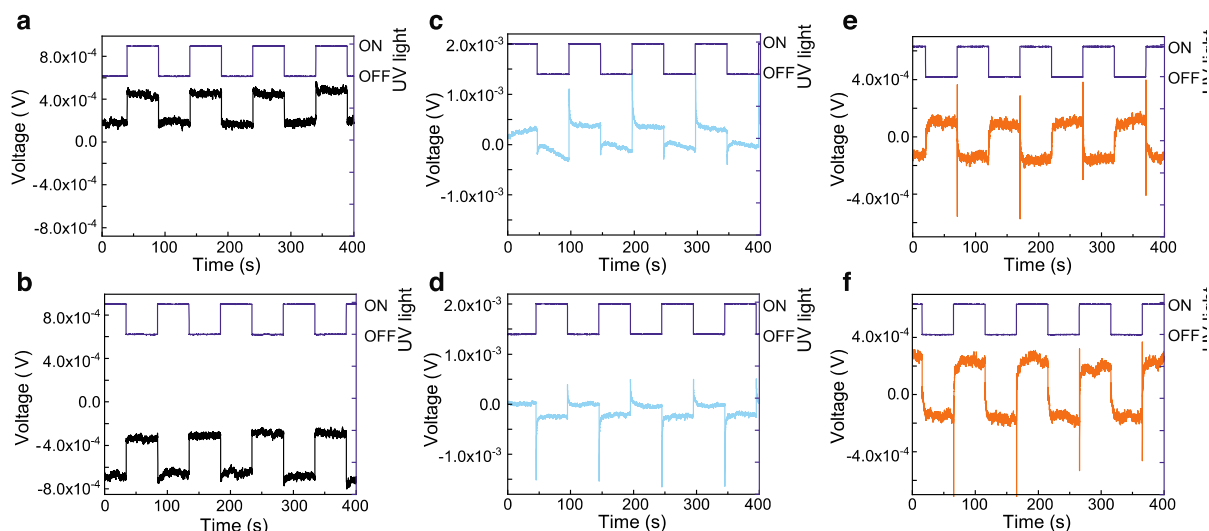

**Supplementary Figure 9.** Voltage difference developed between electrodes when samples are illuminated with UV light in on/off manner with period of 100 seconds (square wave on the top of each figure indicates light on or off). **a,** and **b**, MAPbI$_3$, thickness t=5.68 mm. **c,** and **d**, MAPbCl$_3$, thickness t=3.87 mm. **e,** and **f**, MAPbBr$_3$, thickness t=4.25 mm. In **(b)**, **(d)**, **(f)** samples are rotated by 180° with respect to orientation in **(a)**, **(c)** and **(e)**. Vertical position of the voltage across the samples is arbitrary, but it changes the sign when the sample is flipped. For UV light application see Supplementary Note 2.

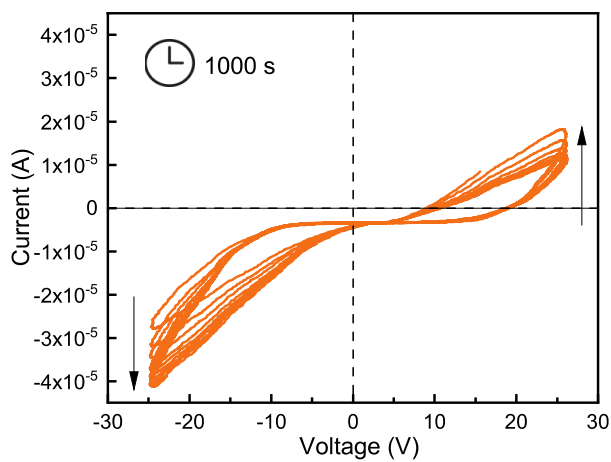

**Supplementary Figure 10.** Asymmetrical current in function of ac voltage (frequency 10 mHz) for an MAPbBr$_3$ sample. Graphite electrodes were sprayed on two parallel faces perpendicular to the growth direction. Thickness of the sample t=3.62mm. Note that the current amplitude and asymmetry evolve with time and become larger over duration of the experiment which lasted 1000 s or ten periods. The arrows indicate direction in which the current evolves.

**Supplementary note 8. Decay of the piezoelectric response after poling**

Supplementary Figure 11 shows decrease of the piezoelectric response amplitude shortly after E$_{dc}$ removal indicating that a substantial fraction of defects population displaced by the poling field tends to relax to their equilibrium position without field. Once this ageing process is finished, in some cases the sample's electro-mechanical response is dominated by the electrostriction.

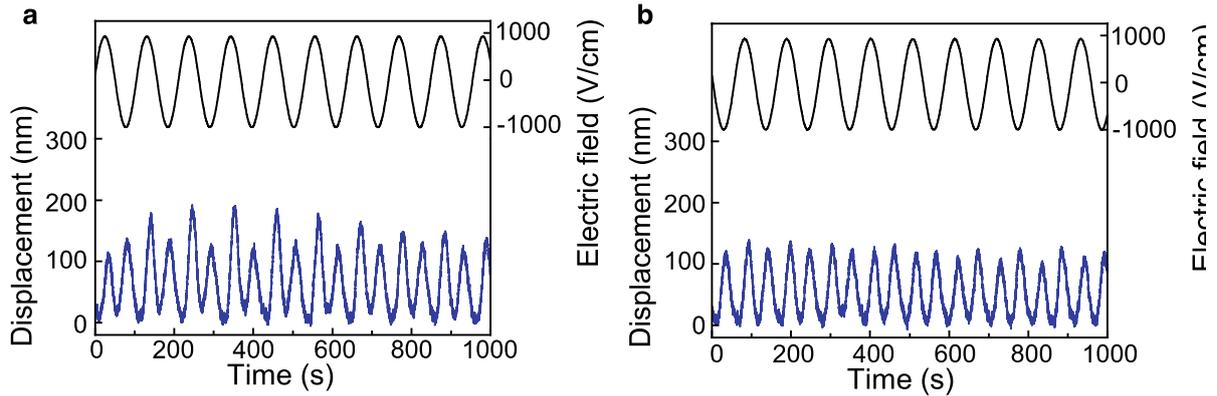

**Supplementary Figure 11.** Electro-mechanical displacement in an MAPbBr$_3$ sample. **a,** Poling voltage of $V_{dc} = -100$ V was first applied for ten minutes, removed, $V_{ac}=150$ V was applied and data were recorded. **b,** 15 minutes after data in **(a)** were taken. The ac voltage was kept applied during the waiting time. The sample had thickness t=1.54 mm and graphite electrodes were sprayed on faces perpendicular to the crystal growth (electric field applied along the growth direction).

**Supplementary Note 9 Evolution of hysteresis with electric field**

As shown in Supplementary Figure 12, the bipolar hysteresis in MAPbBr$_3$ resembles strain-electric field "butterfly loops" in ferroelectric materials[7] where origin of the hysteresis is in switching of domain walls. The behaviour in MAPbX$_3$ could be arguably called ferroionic-like behaviour[8], but with absence of ferroelectricity, at least for X=Cl and Br. The evolution of the hysteresis with electric field indicates that units contributing to the strain have a broad distribution of "coercive" or activation fields (see field dependence of the hysteresis minimum). The hysteresis is not only a consequence of a slow process (time lag). The field dependence indicates that contributing units move in a potential energy landscape with a distribution of barriers. The asymmetry of the loops indicates presence of a bias field. Note also that the loops measured parallel and antiparallel to growth direction are different. This could be caused by internal bias field in the sample. In addition, the state of the sample could have been changed by the first set of measurements.

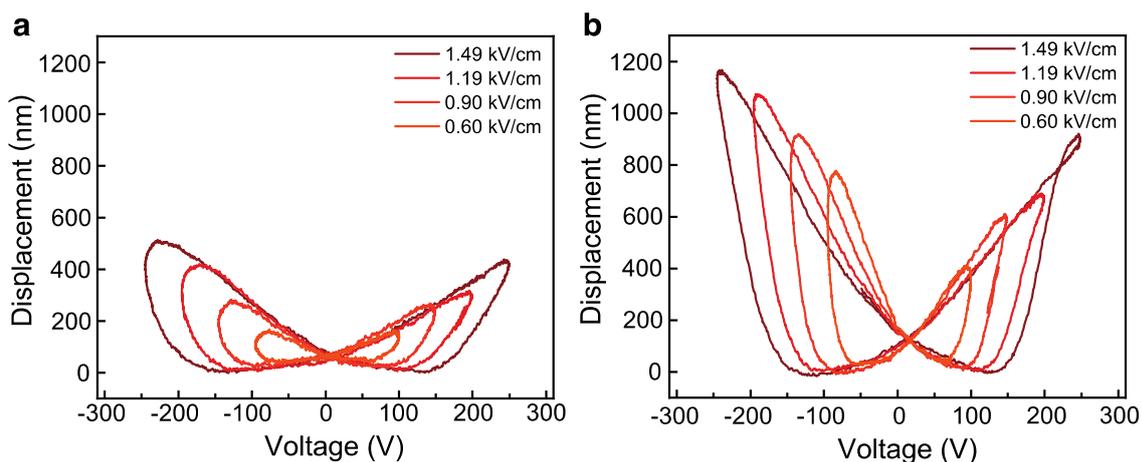

**Supplementary Figure 12.** Displacement versus voltage loops in an MAPbBr$_3$ crystal with sprayed graphite electrodes and thickness t=1.68 mm. **a)** growth direction "down"; **b)** sample flipped over with growth direction "up". Amplitude of the ac field is indicated in each figure. The position of the loops along the vertical axis is arbitrary.

**Supplementary note 10. Frequency dependence of the electro-mechanical response**

An illustration of the frequency dependence of the electrostrictive and piezoelectric coefficients is given in Supplementary Figure 13a for a MAPbCl$_3$ crystal. In the frequency range 10 mHz to 4 Hz, both coefficients decrease monotonously on the log-log scale. The frequency dependence of the dielectric permittivity, which could have been measured over a broader frequency range, is shown in Supplementary Figure 13b for a MAPbBr$_3$ sample. The permittivity shows levelling off above 100 Hz, and reaches values comparable to those reported elsewhere in the literature.[4] Debye type relaxation for the permittivity, with parameters similar to those in experimental data, is shown for illustration to indicate departure of the experimental data from a model of simple Debye relaxation (one kind of contributing units with single activation energy). The departure from the model, which is especially pronounced in the imaginary part of the permittivity, suggests a high level of disorder in the investigated MAPbX$_3$ crystals.[9,10] The disorder could be related to a random energy potential in which defects and molecules move.[11]

The permittivity was measured using a lock-in amplifier (SR 830) and a charge amplifier (Kistler 5011). The output voltage signal from the Lock-in was applied on one electrode and induced charge was collected from the opposite electrode, processed by the charge amplifier and the output signal from the charge amplifier was then fed into the input of the lock-in amplifier and read at the 1$^{st}$ harmonic together with the phase angle between the field and the response. This allowed calculation of charge density developed at the sample and consequently calculation of the complex permittivity of the sample. The frequency of the output signal was swept from 10 kHz to 100 mHz and measurements performed at selected frequencies.

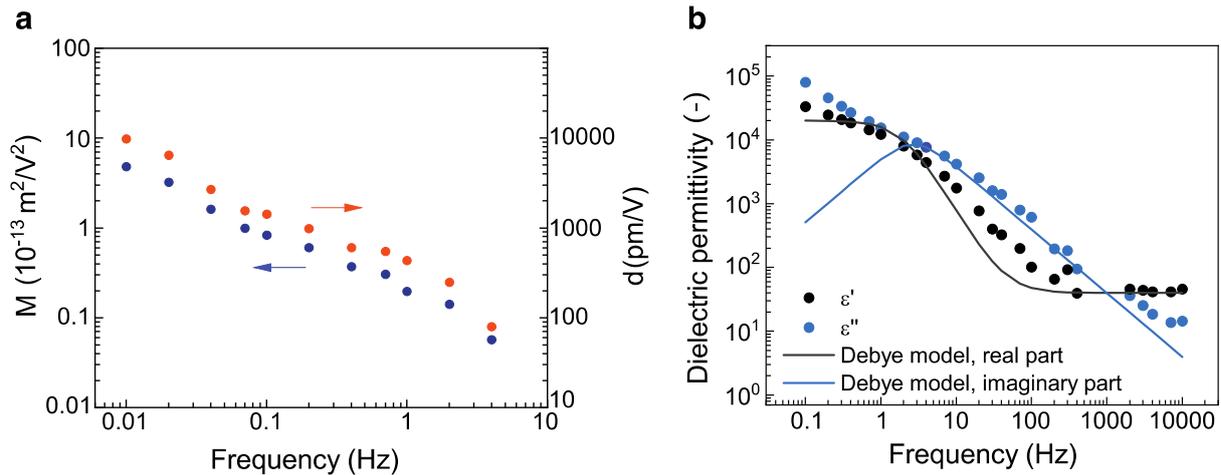

**Supplementary Figure 13. a)** Frequency dependence of M and d coefficients of an MAPbCl$_3$ sample. The thickness t=3.8 mm had sprayed graphite electrodes. Vac=150. **b)** Frequency dependence of the relative complex dielectric permittivity of an MAPbBr$_3$ sample with thickness t=1.68 mm (closed circles). V$_{ac}$=0.7 V. The lines represent relaxation that would be expected for a system with Debye relaxation. Compare frequency scales in **(a)** and **(b)**.

**Supplementary note 11. Change of internal polarization and properties with time**

An important characteristic of the investigated MAPbX$_3$ samples is evolution of their properties with time, while samples are being driven by an external ac field. The time dependence of properties may explain apparent irreproducibility or inconsistency of measurements, often reported in various papers on MAPbX$_3$. This time dependence is caused by substantial role the defects redistribution has in controlling the properties. The effect is particularly strong on the amplitude and even the direction of "forbidden" polarization. Because of the strong sample's sensitivity on external fields every measurement gives only a snapshot of the behaviour at the moment of observations.

Supplementary Figures 14 and 15 give examples of properties evolution with time.

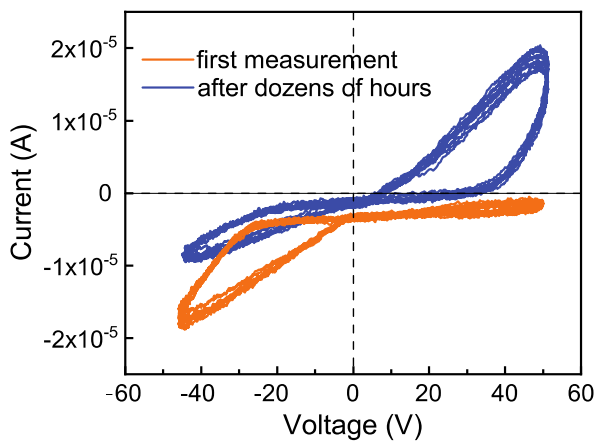

**Supplementary Figure 14.** I-V relation for an MAPbBr$_3$ sample with thickness t=6.69 mm and sprayed graphite electrodes. The voltage was applied with frequency of 10 mHz and current measured on two faces parallel to the growth direction. The current was strongly asymmetrical and was negative at the beginning (orange line) for both negative and positive voltage. After intermittently driving the sample with voltage for dozens of hours, the current asymmetry changed and the current partially changed the direction (blue line).

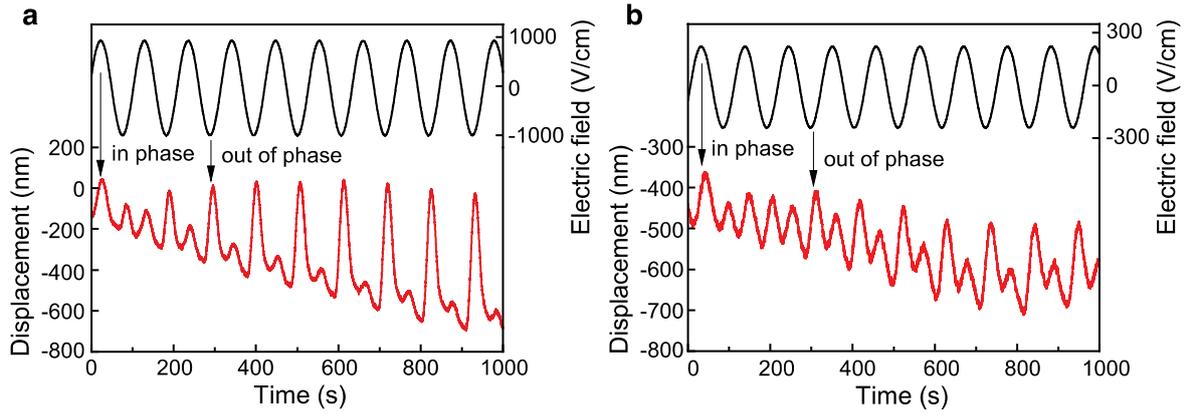

**Supplementary Figure 15.** Change of phase of the electro-mechanical signal during testing of MAPbBr$_3$ samples. Both samples had sprayed graphite electrodes and were first poled with 100 V$_{dc}$ for 10 minutes before making ac measurements. **a,** The field applied along the crystal growth direction, thickness t=1.54 mm. **b,** The field applied perpendicular to the crystal growth direction, thickness t=6.4 mm.

**Supplementary Note 12. Asymmetrical currents, Joule heating and thermal expansion**

A simple model shows that in samples with asymmetrical I-V curves (see Supplementary Figure 10 and 14), the thermal contribution to mechanical displacement (i.e., thermal expansion due to Joule heating) could look similar to electrically excited electrostrictive and piezoelectric displacement.

If a sample is driven by voltage V and the resulting current is I, the power dissipation P is given by $P = IV$. The increase in the temperature $\Delta T$ of the sample because of the Joule heating is given by $\Delta T \sim tP/Cm$ where t is the time, C the heat capacity and m is the mass of the sample. For this simple model, we neglect the heat loss from the sample into environment. The thermal expansion $\Delta l$ is given by $\Delta l \sim \alpha \Delta T$. It follows that $\Delta l \sim \alpha \Delta T \sim \frac{\alpha t P}{Cm}$. For a resistive circuit with $V = V_0 \sin(\omega t)$ and Ohmic I-V relationship, the instantaneous power is $P \sim I_0 V_0 \sin^2(\omega t)$. It follows that the thermally induced mechanical displacement $\Delta l \sim P \sim \sin^2(\omega t)$, will behave as electrostrictive strain (see Supplementary Note 3 and Supplementary Figure 16a).

     If the I-V relation is diode-like (see Supplementary Figure 10 and 14) it may be represented by relation $I = I_0 * (e^{a*V0*\text{Sin}[\omega*t]} - 1)$ where parameter *a* describes the exponential increase of the current. In that case, the power (and Joule heating induced thermal expansion $\Delta l$) will resemble mixed electric-field induced piezoelectric and electrostrictive displacement, see Supplementary Figure 16b and compare with Figure 1a. In fact, this will happen for any asymmetrical I-V relationship.

     In Supplementary Note 13 and the text (Section Thermal expansion) we show that in the samples investigated here the electrically induced displacement is dominated by the electrostriction and piezoelectricity and only a minor part by the electric-field induced thermal expansion through Joule heating of the samples.

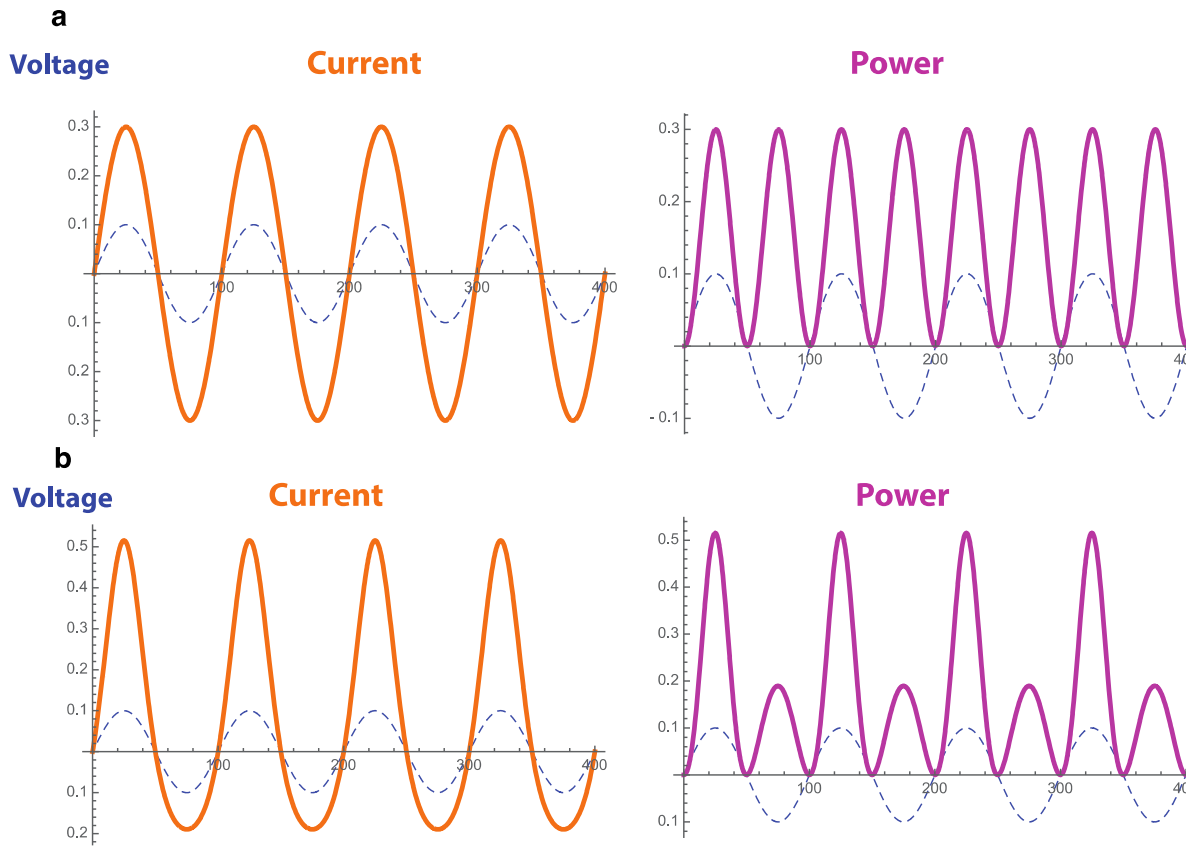

**Supplementary Figure 16. a,** In a resistive circuit with ohmic I-V relationship and V=V$_0$sin(ωt) (dashed blue line), the thermal displacement $\Delta l$ due to Joule heating will resemble pure electrostrictive strain ($\Delta l \sim P \sim \sin^2(\omega t)$). **b,** If I-V relationship is diode-like, (V=V$_0$sin(ωt) and $I = I_0 * \left(e^{a*V0*\text{Sin}[\omega*t]} - 1\right)$), the thermal displacement $\Delta l$ will resemble mixed piezoelectric-electrostrictive strain. In both panels the current is represented with the full orange line and the power with the full purple line.

**Supplementary Note 13. x-ray diffraction**

X-ray diffraction intensity was collected in the form of the rocking curves around 004 Bragg reflection. Specifically, diffraction intensity distribution over the two-dimensional detector was measured at 50 different rocking ("omega") angles covering 0.3 degrees range around the peak position. While staying at each "omega" angle, we collected 100 detector frames each 1 second long, simultaneously applying 100 s long voltage cycle to the crystal. This way full three-dimensional intensity distribution in the reciprocal space was accumulated as a function of time and applied electric field. See Supplementary Figure 17.

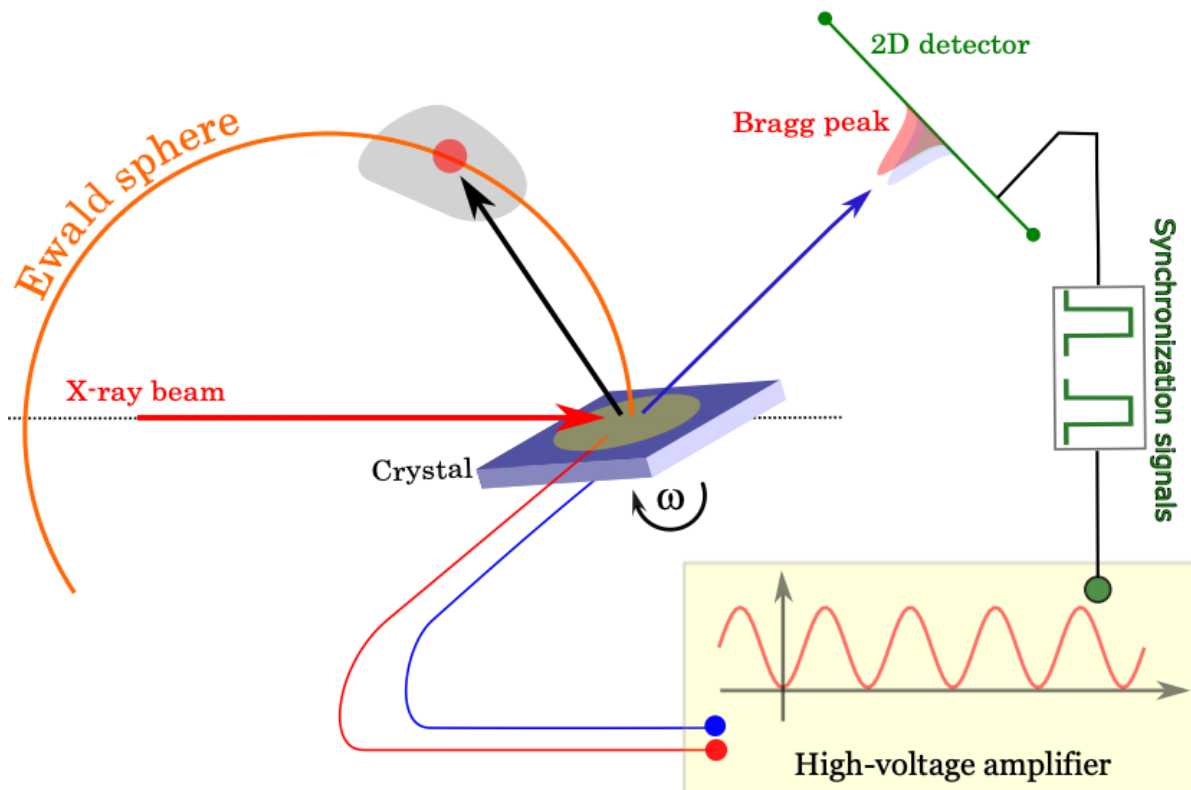

**Supplementary Figure 17.** Schematic of the set-up used for in–situ x-ray diffraction experiment.

Supplementary Figure 18a shows the temperature dependence of 004 rocking curves. From the displacement of the peak position in the reciprocal space we calculate the temperature-driven strain. In addition, we calculated the integrated intensity of the peaks. Supplementary Figure 18b shows the temperature dependence of the strain and the integrated intensity. Note that the temperature-dependent strain manifests the thermal expansion.[1] The decay of the integrated intensity results from the temperature-dependent contribution of the Debye-Waller factor to the structure factor.

     A thermocouple, placed at the contact-free edge of the sample was used for the temperature dependent measurements. In the investigated sample application of ac voltage with negative dc voltage causes the periodic variation of the temperature of the amplitude of 2 K. Positive dc voltage results in negligibly small temperature variations. This difference is due to asymmetry in current between positive and negative voltages (see Supplementary Figures 10 and 14). In this case, ac voltage with positive offset generates very small current and therefore negligible heating of the sample.

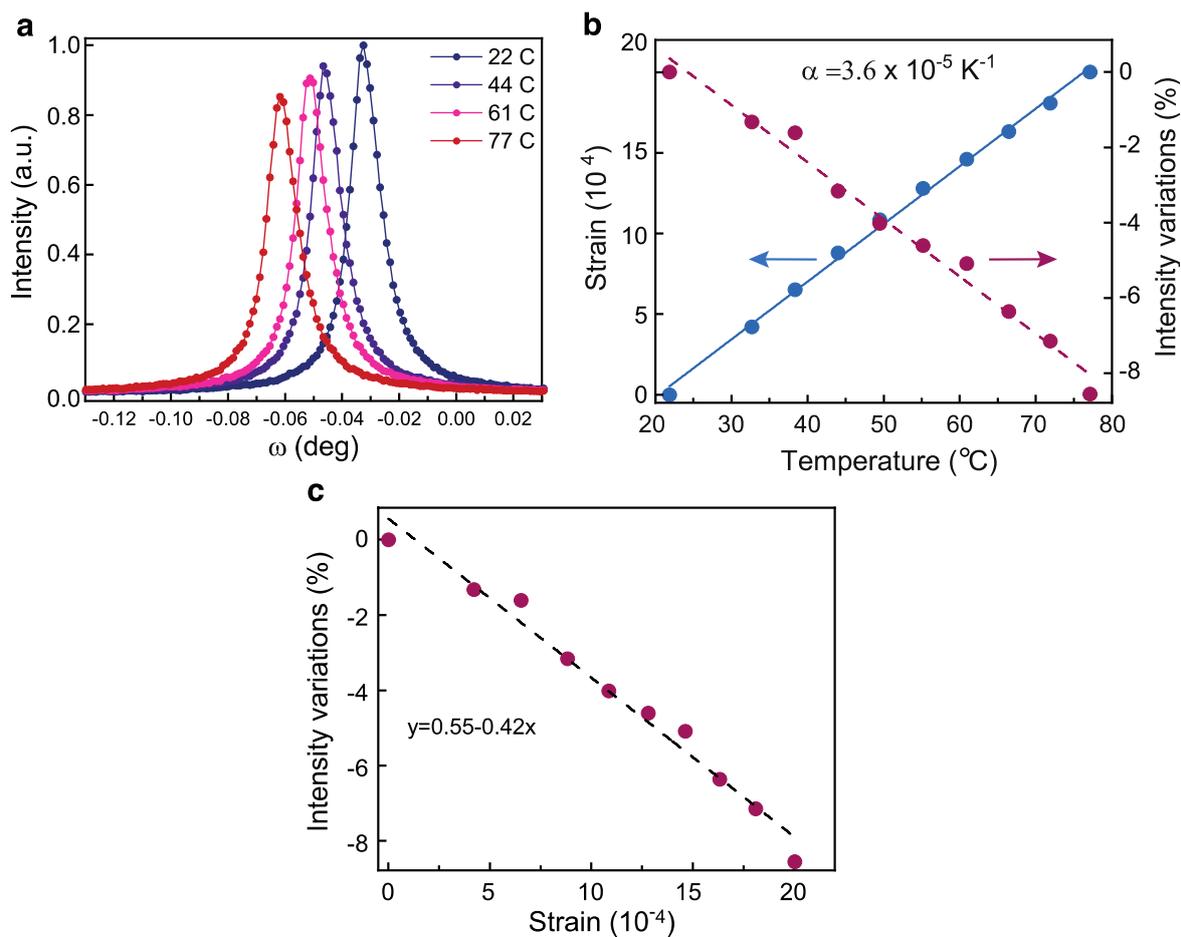

**Supplementary Figure 18.** The measurements of 004 reflection at various temperatures in the range between 20 °C and 80 °C. **a)** the rocking curves of 004 profiles measured at four different temperatures. The measurements were performed at ten temperatures, four of which are retained in the figure for clarity. **b)** The temperature dependence of strain and integrated intensity. **c)** the dependence of integrated intensities on strain.
MAPbBr3, Cs electrodes, t=1.74 mm, AC 100Vpp, DC +50 and -50V

**References:**

1. Ge, C. *et al.* Ultralow Thermal Conductivity and Ultrahigh Thermal Expansion of Single-Crystal Organic–Inorganic Hybrid Perovskite CH3NH3PbX3 (X = Cl, Br, I). *J. Phys. Chem. C* **122**, 15973–15978 (2018).

2. Davis, M. Phase transitions, anisotropy and domain engineering : the piezoelectric properties of relaxor-ferroelectric single crystals. (Swiss Federal Institute of Technology - EPFL, 2006).

3. Morozov, M. Softening and hardening transitions in ferroelectric Pb(Zr,Ti)O3 ceramics. (Ecole polytechnique fédérale de Lausanne, 2005).

4. Onoda-Yamamuro, N., Matsuo, T. & Suga, H. Dielectric study of CH3NH3PbX3 (X = Cl, Br, I). *Journal of Physics and Chemistry of Solids* **53**, 935–939 (1992).

5. Chen, B. *et al.* Large electrostrictive response in lead halide perovskites. *Nature Materials* **17**, 1020 (2018).

6. Newnham, R. E., Sundar, V., Yimnirun, R., Su, J. & Zhang, Q. M. Electrostriction: Nonlinear Electromechanical Coupling in Solid Dielectrics. *The Journal of Physical Chemistry B* **101**, 10141–10150 (1997).

7. Maqbool, A. *et al.* Evolution of ferroelectric and piezoelectric response by heat treatment in pseudocubic BiFeO3–BaTiO3 ceramics. *J Electroceram* **41**, 99–104 (2018).

8. Liu, Y. *et al.* Ferroic Halide Perovskite Optoelectronics. *Advanced Functional Materials* **31**, 2102793 (2021).

9. Jonscher, A. K. The 'universal' dielectric response. *Nature* **267**, 673–679 (1977).

10. Jonscher, A. K. *Dielectric relaxation in solids*. (Chelsea Dielectric Press, 1983).

11. Damjanovic, D. Hysteresis in piezoelectric and ferroelectric materials. in *The Science of Hysteresis* (eds. Bertotti, G. & Mayergoyz, I.) vol. III 337–465 (Academic Press, 2006).